\documentclass{article}

\usepackage{arxiv}

\usepackage[utf8]{inputenc} % allow utf-8 input
\usepackage[T1]{fontenc}    % use 8-bit T1 fonts
\usepackage{hyperref}       % hyperlinks
\usepackage{url}            % simple URL typesetting
\usepackage{booktabs}       % professional-quality tables
\usepackage{amsfonts}       % blackboard math symbols
\usepackage{nicefrac}       % compact symbols for 1/2, etc.
\usepackage{microtype}      % microtypography
\usepackage{cleveref}       % smart cross-referencing
\usepackage{lipsum}         % Can be removed after putting your text content
\usepackage{graphicx}
\usepackage{subcaption}
\usepackage[authoryear]{natbib}
\usepackage{doi}
\usepackage[utf8]{inputenc}
\usepackage[T1]{fontenc}

\DeclareUnicodeCharacter{2212}{\textendash}
\DeclareUnicodeCharacter{251C}{\textendash}
\DeclareUnicodeCharacter{2310}{\textendash}
\DeclareUnicodeCharacter{0393}{\textendash}
\DeclareUnicodeCharacter{03B2}{\ensuremath{\beta}}

\title{Cross-scale spatially-aware generative modeling of transcriptomic programs underlying neurodegenerative brain organization}

\date{}

\newif\ifuniqueAffiliation
\uniqueAffiliationtrue

\ifuniqueAffiliation % Standard variant of author block
\author{ \href{https://orcid.org/0000-0001-5114-9866}{\includegraphics[scale=0.06]{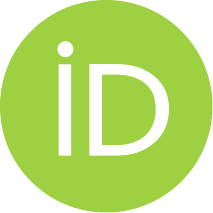}\hspace{1mm}Krishnakumar ~Vaithianathan}\thanks{Corresponding Author} 
	 \\
	Department of Computer Engineering, \\
	Karaikal Polytechnic College, \\
	Karaikal, Puducherry, India.	\\
	\texttt{vkichu77@hgmail.com} \\
	\And
	\textbf{for the Alzheimer's Disease Neuroimaging Initiative}
}
\renewcommand{\shorttitle}{\textit{arXiv} Template}

\hypersetup{
pdftitle={A template for the arxiv style},
pdfsubject={q-bio.NC, q-bio.QM},
pdfauthor={David S.~Hippocampus, Elias D.~Striatum},
pdfkeywords={First keyword, Second keyword, More},
}

\begin{document}
\maketitle

\begin{abstract}
	Neurodegenerative disorders such as Alzheimer’s disease exhibit highly organized patterns of regional brain vulnerability, yet the biological mechanisms underlying this spatial selectivity remain incompletely understood. While previous imaging-transcriptomic studies have primarily focused on correlation-based analyses between gene expression and neuroimaging phenotypes, these approaches often lack generative biological representations capable of modeling how transcriptomic organization gives rise to large-scale neurodegenerative structure. In this study, we introduce a cross-scale spatially-aware generative framework for modeling transcriptomic programs underlying cortical neurodegeneration.
	
	Regional transcriptomic profiles were derived from the Allen Human Brain Atlas using 910 landmark genes aggregated across 68 cortical regions. Neurodegenerative vulnerability maps were constructed from ADNI FreeSurfer cortical thickness measurements by computing regional cortical thinning differences between cognitively normal controls (NC = 926) and Alzheimer’s disease subjects (AD = 426). A variational generative architecture was then used to learn latent biological programs linking regional gene-expression organization to macroscale cortical degeneration. To preserve biologically plausible spatial organization, the model additionally incorporated graph-based spatial smoothness regularization across neighboring cortical regions.
	
	The proposed framework achieved strong prediction of regional neurodegenerative vulnerability, yielding an $R^2$ score of $0.8604$ and a significant spatial correlation of $r=0.9439 (p<0.001)$ between predicted and observed cortical degeneration profiles. The learned latent representations further revealed structured transcriptomic organization associated with spatially distributed neurodegenerative susceptibility.
	
	Together, these findings demonstrate that biologically constrained generative modeling can bridge microscale transcriptomic organization with macroscale neurodegenerative brain structure. This work provides a foundation for spatially-aware generative neurobiology and offers a scalable framework for studying cross-scale mechanisms underlying neurodegenerative disease organization.
\end{abstract}

% keywords can be removed

\section{Introduction}
\subsection{Spatial Neurodegeneration}
Neurodegenerative disorders do not take place at random locations in the brain. In contrast, the spatial organization of pathological processes is highly structured in large scale brain systems in the context of neurodegeneration \cite{seeley2009neurodegenerative,raj2012network}. In Alzheimer's disease (AD), temporoparietal association cortices, limbic structures and widespread transmodal networks undergo predictable degeneration of the cortex \cite{braak1991neuropathological,thompson2003dynamics}. The patterns indicate that local pathology isn't solely responsible for regional vulnerability, but that other intrinsic biological and network-level characteristics also play a role.

In recent years, there have been several studies demonstrating how the architecture of the human connectome is associated with the spread of disease \cite{zhou2012predicting,vogel2020spread}. Structurally and/or functionally interconnected regions tend to exhibit correlated vulnerability, as it is theorized that trans-neuronal spread and network diffusion propagate vulnerability \cite{raj2015network}. This view has significantly shifted the focus of computational neuroscience from a voxel-wise description to systems-level models of degeneration. It is therefore becoming a newer paradigm for research to consider AD as a disorder of spatially embedded brain organization.

Meanwhile, regions susceptibility will not be entirely accounted for by connectivity. Another important correlate of the different rates of degeneration in different regions of the brain seems to be molecular heterogeneity within the brain's various "cortical territories" that differ in their genetic composition \cite{hawrylycz2012anatomically}. Macroscale disease organization may then have a biological basis in terms of spatial variations in transcriptional architecture. This concept has inspired recent studies that investigated the connection of transcriptomics with neuroimaging-based phenotypes, including those related to cortical thinning, accumulation of amyloid, propagation of tau, and susceptibility of the network \cite{fornito2019bridging,mroczek2021imaging}.

Importantly, there is simultaneous multiple biological scale-longitudinal progression of cortical degeneration. The molecular at the micro-scale influences the organization at the meso-scale (the level of the cell) and this organization at the meso-scale influences the change of the neuro-anatomical structure at the macro-scale. One of the major problems in computational neurobiology is modelling these interactions. Current statistical methods are generally not well-suited to describe the hierarchical biological organization, which alsojustifies the development of approaches that are more generative and spatially-aware, and which are able to connect the transcriptomic programs with the degeneration pattern of systems.

Recently, spatial biological modeling has gained even more significance with emerging spatial transcriptomics and AI-powered neuroscience techniques such as \cite{zhao2024spatial, murray2025accelerating}. A growing number of emerging studies highlight the importance of understanding neurodegeneration as a spatial process of coordinated gradients of transcriptomics, connectomics and the landscapes of vulnerability unique to each disease.

\subsection{Imaging Transcriptomics}
Imaging transcriptomics has evolved to become a powerful tool for associating molecular biology with the macroscale organization of the brain. Connecting regional gene expression atlas data with neuroimaging phenotypes allows researchers to explore the variation of brain structural and functional properties that result from different patterns of gene expression across the brain \cite{fornito2019bridging,arnatkeviciute2023toward}. The Allen Human Brain Atlas (AHBA) has emerged as a key resource to this field as it offers a given set of anatomically resolved transcriptomic measurements throughout the adult human cortex \cite{hawrylycz2012anatomically}.

In the last decade, imaging transcriptomics studies have linked molecular signatures to different processes of the brain, including developmental gradients, neurodegenerative vulnerability, neurotransmitter systems, and functional connectivity, as well as cortical thickness \cite{mroczek2021imaging,adewale2021integrated}. For Alzheimer's disease, however, a number of studies have shown that regions with transcriptional connectivity show associations with either amyloid load, tau load or atrophy in the cortex \cite{brusini2024morphometric} among other things. The findings indicate that this organization of the genome, transcriptomes, may contribute to selective susceptibility to degeneration of some cortical systems.

Despite the developments, the vast majority of imaging transcriptomic studies are still correlative. Typical analysis methods include spatial correlation analysis, enrichment testing of imaging-derived maps and gene expression profiles or partial least squares regression (PLS) analysis of the maps \cite{arnatkeviciute2023toward}. These approaches are informative, but are generally not capable of modelling latent biological processes or of nonlinear cross scale interactions. As a result, many current methods give description based associations rather than the biologically meaningful generative representations.

A series of recent advances in machine learning have started to change all that. The deep generative models such as variational autoencoders (VAEs) and generative adversarial networks (GANs) have the potential to learn compressed latent structures that capture the hidden biological organization \cite{pinaya2022brain,kingma2014autoencoding,ramezanian2022generative}. These approaches are especially appealing for computational neuroscience, where they can encode complex relationships of high dimensional molecular information with brain phenotypes that are nonlinear.

Hereby, in the era of AI-empowered neurogenomics, multimodal integration, graph learning, and spatial priors are becoming more and more common in the field of neurogenomics research (e.g., \cite{liu2025radiogenomics, interpretable2025genomics}). More recently, studies on generative frameworks for spatial transcriptomics and disease modelling have emerged in the fields of medicine and neuroscience in particular, such as in \cite{yu2025tissue,zhang2025brainbeacon}. These developments suggest a general paradigm shift in association studies to biologically informed generative modelling, which can depict cross-scale neurobiological organization.

At the same time, large-scale neuroimaging efforts like the Alzheimer's Disease Neuroimaging Initiative (ADNI) have allowed to define well disease-related degeneration of the cortex in thousands participants \cite{jack2008alzheimer}. When coupled with image resources and transcriptomic atlases, this could be a unique opportunity to test whether molecular programs are latent that could drive spatiotemporal patterns of neurodegeneration.

\subsection{Existing Limitations}
Imaging transcriptomics has proven to be a valuable tool for understanding the organization of brains, but there are some methodological issues that remain poorly solved. To date, a great number of studies use pairwise correlation analysis between regional imaging phenotypes with gene expression measurements \cite{arnatkeviciute2023toward}. Such strategies uncover statistical relationships, but do not readily reveal hierarchical biological interactions or hidden molecular programs that could be involved in disease organization.

Secondly, existing structures are not generative. Traditional linear-based models and enrichment analysis usually do not learn compressed representations of the imaging and transcriptomic structure, but rather map the genes directly to the imaging features \cite{mroczek2021imaging}. Consequently, these approaches might neglect such nonlinear biological dependencies, which can span many genes and many cortical systems. Such simplifications can greatly reduce biological interpretability, however, in neurodegenerative disorders like Alzheimer's disease, where the progression of the disease is the result of interacting molecular and network-level mechanisms.

Thirdly, lots of methods do not consider spatial relationships in neighboring cortical regions. Brain organization is always constrained in space, and neighboring regions often have highly correlated structural, molecular and functional features \cite{huntenburg2018large,margulies2016situating}. But traditional machine learning approaches tend to assume that brain regions are independent samples, ignoring the topological relationships they have amongst each other in the brain's anatomy. One of the limitations is that while neurodegeneration in AD occurs on a spatially organized pathway, rather than spatially isolated events, as in other diseases, the extent to which the processes of ordered versus spatially-isolated degeneration influence development remains unclear.

The other major concern is regarding the scale-integration of biology. Studying the transcriptomic organization or neuroimaging phenotypes independently, without making explicit models linking microscale molecular systems with macroscale disease vulnerability is common in existing studies. Thus, there is still a big gap between systems neuroscience and molecular neurobiology. The goal of creating computational frameworks to understand these scales is now seen as one of the main objectives of modern neuroscience research efforts \cite{murray2025accelerating,cardillo2025multiomics}.

In recent years, AI-driven solutions have come to the rescue with multimodal deep learning and graph-based methods \cite{liu2025radiogenomics}, which have begun to tackle some of these challenges. However, the majority of existing approaches still focus on getting the predictions right instead of being interpretable from a biological angle. Moreover, relatively few studies incorporate biologically constrained latent representations together with explicit spatial regularization mechanisms. This leaves an important opportunity for computational models that integrate transcriptomic encoding, latent biological programs, and spatially-aware cortical organization within a unified generative framework.

In this study, we address these limitations by developing a cross-scale spatially-aware generative model linking regional transcriptomic organization with neurodegenerative brain vulnerability. Rather than treating gene expression and cortical degeneration as independent modalities, our framework learns biologically constrained latent representations capable of predicting regional neurodegeneration while preserving spatial smoothness across cortical systems.

\subsection{Cross-Scale Spatially-Aware Generative Neurobiology}
To address the limitations of existing imaging transcriptomics approaches, we propose a cross-scale spatially-aware generative framework that models the relationship between regional transcriptomic organization and neurodegenerative vulnerability. The central objective of the model is to learn biologically meaningful latent programs capable of explaining how microscale molecular architecture contributes to macroscale cortical degeneration patterns in Alzheimer's disease.

Our framework is motivated by the observation that cortical neurodegeneration emerges through coordinated interactions across multiple biological scales. Regional gene expression profiles influence cellular identity, synaptic organization, metabolic regulation, and network-level communication, all of which contribute to regional disease susceptibility \cite{fornito2019bridging,murray2025accelerating}. Consequently, modeling neurodegeneration requires computational architectures capable of integrating high-dimensional molecular information with spatially organized brain phenotypes.

The proposed model operates on region-level transcriptomic measurements derived from the Allen Human Brain Atlas (AHBA) \cite{hawrylycz2012anatomically}. After filtering to landmark genes from the LINCS L1000 platform \cite{subramanian2017next}, transcriptomic profiles are aligned with cortical neurodegeneration maps derived from FreeSurfer-based cortical thickness measurements in the Alzheimer's Disease Neuroimaging Initiative (ADNI) cohort \cite{jack2008alzheimer}. Regional neurodegeneration vulnerability is quantified as the difference between cognitively normal and Alzheimer's disease cortical thickness profiles across anatomically matched cortical regions.

Unlike conventional regression-based imaging transcriptomics approaches, our framework adopts a generative representation learning strategy. Specifically, we employ a variational autoencoder (VAE) architecture to encode regional transcriptomic profiles into low-dimensional latent representations \cite{kingma2014autoencoding,doersch2016tutorial}. These latent variables are interpreted as transcriptomic programs that capture hidden biological organization across cortical systems. The decoder infers regional gene expression profiles based on the latent embeddings, guaranteeing the learned representations being biologically meaningful molecular structure.

The second component of the network projects latent transcriptomic representations onto cortical neurodegeneration vulnerability. This design facilitates the model to learn the particular non-linear relationships between the transcriptomic organisation and the patterns of cortical thinning associated with the disease. Importantly, the framework is not about assigning some diagnostic label but rather defines the regional phenotype of neuro-degeneration susceptibility as a continuous susceptibility process. This distinction greatly enhances the biological interpretability, since the degeneration of the cortex is a graded phenomenon and is also regionally variable.

Next, spatially regularized learning is added to the biologically plausible organization of the cortex. Molecular, structural and functional characteristics are often layered with correlated characteristics in neighbouring cortical regions, because of shared molecular, developmental and connectomic constraints among them \cite{huntenburg2018large,margulies2016situating}. Thus, model predictions should be smooth in space, and not have anatomically unrealistic discontinuities. We enforce this constraint through graph-based smoothness regularisation, which is based on the methods found in geometric deep learning literature \cite{li2018deeper,cao2020comprehensive}. The resulting objective encourages adjacent cortical regions to maintain similar vulnerability estimates while still allowing disease-relevant heterogeneity to emerge.

This cross-scale formulation differs fundamentally from traditional imaging transcriptomics studies. Rather than computing direct gene-to-imaging correlations, the proposed framework learns latent biological representations capable of generating spatial neurodegeneration organization. In this sense, the model represents a form of generative neurobiology in which hidden transcriptomic programs are used to explain disease-related cortical vulnerability patterns.

Recent advances in AI-driven neurogenomics further support this direction. Emerging multimodal frameworks increasingly combine transcriptomics, neuroimaging, and generative machine learning to model biological systems at scale \cite{liu2025radiogenomics,interpretable2025genomics}. Similarly, foundation-model approaches in computational biology and spatial transcriptomics demonstrate the growing importance of latent representation learning for complex biological organization \cite{tejada2025nicheformer,zhang2025brainbeacon}. Our framework extends these ideas into the context of spatial neurodegeneration by integrating transcriptomic encoding, generative modeling, and cortical spatial constraints within a unified computational architecture.

\subsection{Study Objectives and Contributions}
The primary objective of this study is to develop a biologically grounded generative framework capable of linking regional transcriptomic organization with cortical neurodegenerative vulnerability in Alzheimer's disease. More specifically, we aim to determine whether latent transcriptomic programs learned from AHBA gene expression profiles can predict spatial patterns of cortical degeneration observed in ADNI-derived neuroimaging data.

To achieve this objective, we construct a region-level imaging transcriptomics pipeline integrating three major components: transcriptomic representation learning, neurodegeneration vulnerability modeling, and spatial regularization. First, we generate anatomically aligned cortical gene expression profiles using AHBA data mapped onto the Desikan-Killiany atlas \cite{desikan2006automated}. We then restrict the transcriptomic space to landmark genes from the LINCS L1000 platform to improve biological robustness and computational tractability \cite{subramanian2017next}. At the same time, FreeSurfer outputs produce ADNI cortical thickness maps which are used to create regional neurodegeneration effect maps illustrating the thinning of the cortex that is associated with Alzheimer's Disease.

The proposed framework proposes several methodological contributions to computational neuroscience and imaging transcriptomics.

The first is to propose an explicit model paradigm, bridging the micros- and macros- scale, that explicitly connects the organization of molecules at a micros- scale with the structure of the brain at a macros- scale, in neurodegenerative disorders. Most published studies focus on either the transcriptomic or imaging data, while we use a framework that links the two through biologically constrained latent representations .Our approach is different from most studies that study the transcriptomics and imaging data independently, as we use biologically constrained representations to link the data together from both domains.

Secondly, we introduce a novel imaging transcriptomics generative modelling approach. The present research is conducted mainly based on the correlation analysis or linear statistical mapping \cite{arnatkeviciute2023toward, mroczek2021imaging}. In our model, however, latent transcriptomic programs are learned to reconstruct cortex molecular organization and the vulnerability of the cortex is predicted. This identify the hidden biological structure that are not observed in the direct association analysis alone.

Third, it involves explicitly spatial regularization of the learning process. Biologically realistic models preserve spatial smoothness of neighbouring cortical areas which are known to have continuous spatial gradients in organisation and network constraints \cite{margulies2016situating, huntenburg2018large}. The proposed architecture, that integrates spatial constraints provided by the graph, predicts and generates anatomically coherent predictions of neurodegeneration based on known principles of cortical organization.

Fourth, this work can be considered as part of the emerging trend of generative neurobiology that aims at modelling biological systems by learning latent representations and multimodal AI models \cite{ramezanian2022generative,tejada2025nicheformer}. Here, we want to learn biologically meaningful representations for classification and prediction, incorporating the organization of diseases at scales.

Finally, the proposed framework offers a computational foundation for future multimodal extensions concerning neurobiology, spatial transcriptomics and connectomics, as well as for long-term disease modeling. The present study focuses on the vulnerability of the Alzheimer's disease cortex, but the framework of the present study is applicable to other neurodegenerative disorders and spatial neuroscience applications.

The overall goal of this work is to provide a spatially-aware and biologically-informed computational framework to understand the influence of transcriptomic organisation on large-scale structure of the brain in neurodegenerative diseases.

$\:\:\:\:\:\:\:\:\:\:\:\:\:\:\:\:\:\:\:\:\:\:\:$The remainder of this article is organized as follows. Section~\ref{sect:matmeth} describes the datasets, transcriptomic preprocessing pipeline, neurodegeneration vulnerability mapping, atlas alignment procedures, and the proposed cross-scale spatially-aware generative framework. Section~\ref{sect:results} presents the experimental results, including regional neurodegeneration organization, transcriptomic structure, latent biological programs, predictive performance analyses, and comparisons with baseline models. Section~\ref{sect:discuss} discusses the neurobiological implications of the findings, positioning the framework within imaging transcriptomics and computational neuroscience literature while outlining current limitations and future research directions. Finally, Section~\ref{sect:conclu} summarizes the major contributions and highlights opportunities for future development of spatially-aware generative neurobiology frameworks.

\section{Materials and Methods}
\label{sect:matmeth}

\subsection{Datasets}

\subsubsection{Allen Human Brain Atlas (AHBA)}

Regional transcriptomic data were obtained from the Allen Human Brain Atlas (AHBA), a large-scale postmortem transcriptomic resource containing anatomically resolved gene expression measurements across the adult human brain \cite{hawrylycz2012anatomically}. The AHBA consists of microarray-based gene expression profiles collected from six neurotypical adult donor brains. Tissue samples were spatially mapped to standardized neuroanatomical coordinates, enabling integration with neuroimaging-derived cortical parcellations.

To construct a region-level transcriptomic representation suitable for imaging transcriptomics analysis, cortical samples were mapped onto the Desikan-Killiany cortical atlas \cite{desikan2006automated} using the \texttt{abagen} toolbox \cite{markello2021standardizing}. The Desikan-Killiany atlas was selected because it provides anatomically interpretable cortical regions that are widely used in neuroimaging and neurodegeneration studies. Following atlas mapping and regional aggregation, transcriptomic measurements were represented across 68 cortical regions corresponding to bilateral cortical parcels.

Because the AHBA contains expression measurements for thousands of genes, dimensionality reduction at the biological feature level was performed using the LINCS L1000 landmark gene set \cite{subramanian2017next}. Landmark genes were selected because they represent biologically informative transcriptional programs while reducing redundancy within the transcriptomic space. After filtering, the final transcriptomic matrix consisted of 910 landmark genes measured across 68 cortical regions.

The resulting AHBA matrix can therefore be represented as:

\[
X \in \mathbb{R}^{68 \times 910}
\]

where each row corresponds to a cortical region and each column corresponds to a landmark gene expression profile.

\subsubsection{Alzheimer's Disease Neuroimaging Initiative (ADNI)}

Neuroimaging-derived cortical degeneration measurements were obtained from the Alzheimer's Disease Neuroimaging Initiative (ADNI) database \cite{jack2008alzheimer}. Specifically, FreeSurfer-derived morphometric outputs from the UCSFFSX7 dataset were used to characterize regional cortical thickness across cognitively normal (NC) individuals and patients with Alzheimer's disease (AD).

The ADNI cohort included subjects spanning multiple ADNI phases (ADNI1, ADNIGO, ADNI2, and ADNI3). Diagnostic categories were defined using the provided clinical diagnosis labels. For the present study, only cognitively normal controls and Alzheimer's disease subjects were included in the primary analysis in order to model cortical neurodegeneration vulnerability associated with established disease pathology. Mild cognitive impairment (MCI) participants were excluded from the current framework to maintain a clearer disease contrast, although future extensions of the model may incorporate progressive disease staging and multiclass neurodegenerative trajectories.

Regional cortical measurements were extracted from FreeSurfer Thickness Average (TA) variables corresponding to Desikan-Killiany cortical regions. Thickness Average metrics were selected because cortical thinning is a well-established biomarker of neurodegeneration in Alzheimer's disease \cite{thompson2003dynamics}. After anatomical harmonization, 68 bilateral cortical thickness measurements were retained for downstream analysis.

The final cohort used in this study is given in Table.\ref{tab:finalcohort}.
\begin{table}[h]
	\centering
	\caption{ADNI cohort composition used for neurodegeneration modeling.}
	\renewcommand{\arraystretch}{1.2} 
	\label{tab:finalcohort}
	\begin{tabular}{lc}
		\hline
		Group & Number of subjects \\
		\hline
		Cognitively Normal (NC) & 926 \\
		Alzheimer's Disease (AD) & 426 \\
		\hline
		Total & 1352 \\
		\hline
	\end{tabular}
	\renewcommand{\arraystretch}{1} 
\end{table}

Regional cortical thickness values were subsequently aggregated to generate group-level neurodegeneration profiles used as target variables for the generative modeling framework.

\subsection{Transcriptomic Preprocessing}
\begin{figure}[htbp]
	\centering
	\includegraphics[width=\textwidth]{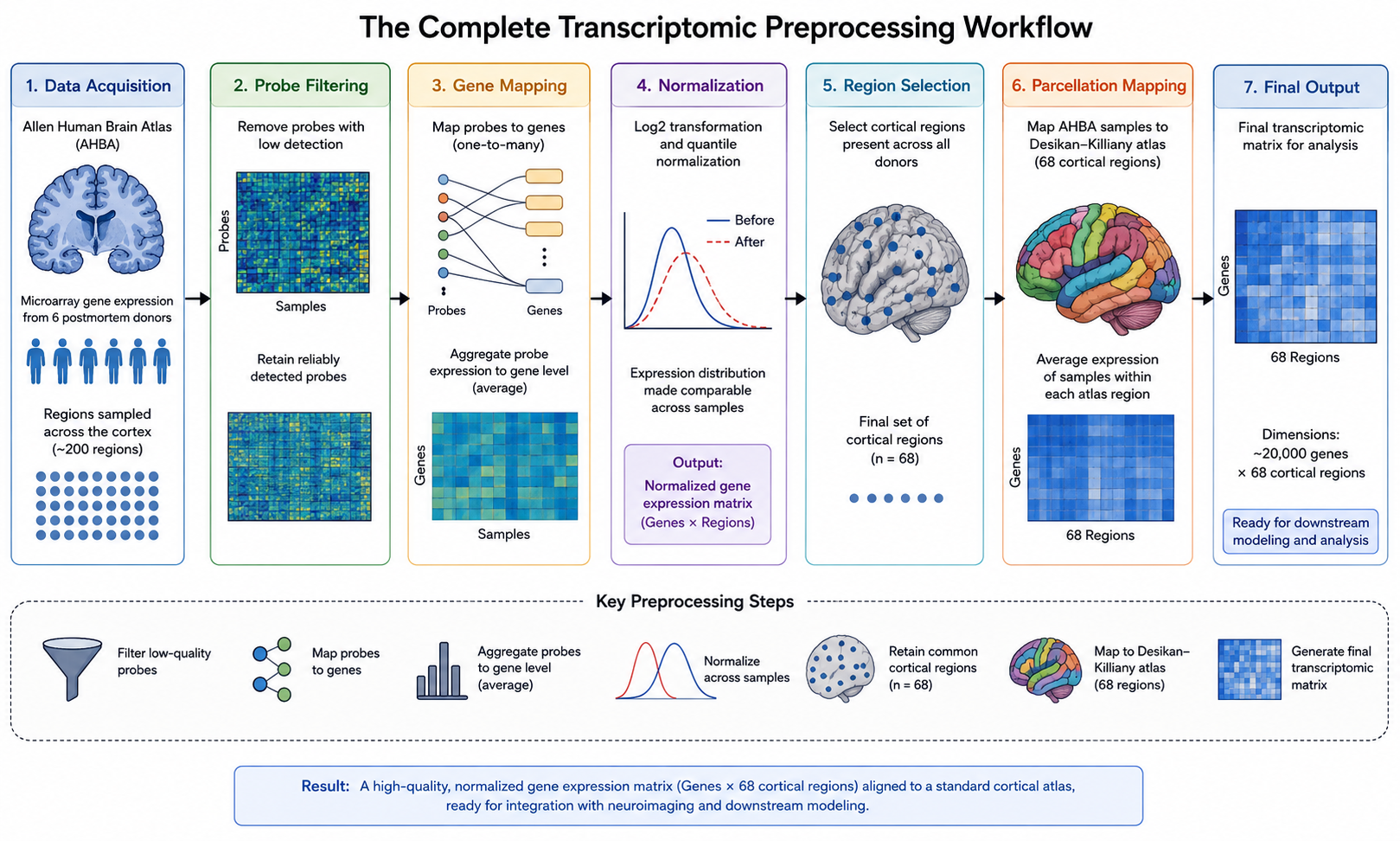}
	\caption{
		Overview of the transcriptomic preprocessing workflow used in the present study. Regional gene expression profiles were obtained from the Allen Human Brain Atlas (AHBA) and processed using the \texttt{abagen} toolbox. Transcriptomic samples were mapped onto the Desikan--Killiany cortical atlas, normalized across donors, and aggregated into region-level cortical expression profiles. LINCS landmark gene filtering was subsequently applied to retain biologically informative transcriptomic features, reducing the initial \(\sim 20{,}000\)-gene cortical transcriptomic matrix to 910 landmark genes across 68 cortical regions for downstream cross-scale generative modeling.
	}
	\label{fig:ahba_pipeline}
	\end{figure}
Transcriptomic preprocessing and atlas mapping were performed using the \texttt{abagen} toolbox, a standardized framework for imaging transcriptomics workflows \cite{markello2021standardizing}. The primary objective of preprocessing was to generate biologically interpretable region-level cortical transcriptomic profiles aligned with neuroimaging-derived cortical regions.

AHBA tissue samples were first assigned to Desikan-Killiany cortical parcels using spatial atlas matching procedures implemented in \texttt{abagen}. To improve hemispheric coverage and reduce sampling imbalance across donor brains, bidirectional hemisphere mirroring was enabled during sample assignment. This approach allows transcriptomic information from homologous cortical regions to contribute to regional expression estimation when direct sampling is limited.

Missing tissue assignments were addressed using interpolation-based procedures available within the \texttt{abagen} framework. Specifically, missing regional values were estimated using neighboring spatial transcriptomic information, improving cortical coverage consistency across atlas regions. A spatial tolerance parameter of 2 mm was used during tissue-to-region assignment to accommodate anatomical variability across donor samples.

Following tissue assignment, regional transcriptomic measurements were aggregated across donors to generate a single region-level expression profile for each cortical parcel. Gene expression values were subsequently normalized according to standard \texttt{abagen} preprocessing recommendations \cite{markello2021standardizing}. These procedures reduce inter-donor variability while preserving biologically meaningful regional transcriptional gradients.

To reduce transcriptomic dimensionality while retaining biologically informative signals, the resulting AHBA matrix was filtered using the LINCS L1000 landmark gene set \cite{subramanian2017next}. Landmark genes are experimentally selected genes designed to capture a substantial proportion of transcriptomic variation while minimizing redundancy. This filtering step improves computational tractability for deep generative modeling while preserving key molecular signatures relevant to cortical organization and disease vulnerability.

After preprocessing and landmark filtering, the final transcriptomic dataset consisted of 68 cortical regions and 910 landmark genes. Region names were harmonized across AHBA and ADNI datasets through standardized formatting procedures involving hemisphere normalization, lowercase conversion, and atlas-consistent anatomical naming. These preprocessing steps enabled direct alignment between transcriptomic and neuroimaging-derived cortical representations.

The complete transcriptomic preprocessing workflow is illustrated in Figure~\ref{fig:ahba_pipeline}. Briefly, the pipeline consisted of atlas mapping, regional aggregation, missing-value interpolation, landmark gene filtering, and cortical region harmonization prior to integration with neurodegeneration vulnerability maps.

\subsection{Neurodegeneration Vulnerability Mapping}

To model regional neurodegenerative susceptibility, a cortical vulnerability map was constructed from ADNI-derived cortical thickness measurements. Regional cortical thickness is a widely used biomarker of Alzheimer's disease progression because cortical thinning reflects neuronal loss, synaptic degeneration, and large-scale structural disruption associated with disease pathology \cite{thompson2003dynamics,jack2018nia}.

For each cortical region, mean cortical thickness values were separately computed for cognitively normal (NC) participants and Alzheimer's disease (AD) participants using anatomically matched Desikan-Killiany cortical regions \cite{desikan2006automated}. Regional neurodegeneration vulnerability was then quantified as the difference between the NC and AD group means:

\[
Y_r = \mu_{NC,r} - \mu_{AD,r}
\]

where:
\begin{itemize}
	\item \(Y_r\) denotes the neurodegeneration vulnerability score for cortical region \(r\),
	\item \(\mu_{NC,r}\) represents the mean cortical thickness in cognitively normal subjects,
	\item \(\mu_{AD,r}\) represents the mean cortical thickness in Alzheimer's disease subjects.
\end{itemize}

Larger values of \(Y_r\) therefore indicate greater cortical thinning in Alzheimer's disease relative to healthy controls and are interpreted as higher regional neurodegenerative vulnerability.

This formulation produces a continuous region-level vulnerability profile rather than a binary diagnostic classification target. Such an approach is biologically advantageous because Alzheimer's disease pathology unfolds gradually across cortical systems and exhibits substantial spatial heterogeneity \cite{seeley2009neurodegenerative,raj2012network}. Modeling regional vulnerability as a continuous spatial phenotype enables the framework to capture graded neurodegeneration patterns that more closely reflect underlying disease biology.

The resulting vulnerability map can be represented as:

\[
Y \in \mathbb{R}^{68 \times 1}
\]

where each entry corresponds to the estimated neurodegeneration vulnerability of a cortical region.

This neurodegeneration representation aligns conceptually with contemporary network-based and spatial models of Alzheimer's disease progression \cite{zhou2012predicting,vogel2020spread}. Rather than treating cortical degeneration as anatomically independent events, the vulnerability framework captures disease-associated spatial organization across distributed cortical systems.

In addition, cortical thickness measurements derived from FreeSurfer provide anatomically interpretable estimates of cortical structural integrity and have been extensively validated in neurodegenerative neuroimaging research \cite{fischl2012freesurfer}. The use of regional cortical thinning profiles therefore enables biologically meaningful integration between transcriptomic organization and disease-related neuroanatomical alteration.

The final vulnerability map served as the target phenotype for the proposed generative neurobiology framework. Specifically, the model was trained to predict spatial neurodegeneration organization from latent transcriptomic representations learned from AHBA gene expression profiles.

\subsection{Atlas Alignment}

Accurate integration of transcriptomic and neuroimaging datasets requires consistent anatomical correspondence across modalities. Because the AHBA transcriptomic data and ADNI cortical thickness measurements originate from independent acquisition pipelines, an atlas harmonization procedure was performed to ensure region-level alignment between datasets.

Both transcriptomic and neuroimaging features were mapped onto the Desikan-Killiany cortical atlas \cite{desikan2006automated}. This atlas provides anatomically interpretable cortical parcellations that are widely used in imaging transcriptomics and neurodegeneration studies \cite{fornito2019bridging,mroczek2021imaging}. Using a common atlas substantially reduces anatomical ambiguity and facilitates biologically meaningful cross-modal integration.

Despite shared atlas usage, minor inconsistencies in cortical region naming conventions remained between the AHBA-derived transcriptomic profiles and the ADNI-derived cortical thickness measurements. To address this issue, region names were standardized using automated harmonization procedures including lowercase normalization, removal of formatting inconsistencies, and hemisphere-consistent anatomical naming.

After standardization, shared cortical regions were identified through direct intersection of transcriptomic and neuroimaging region labels. Only anatomically matched cortical regions present in both datasets were retained for downstream modeling. This procedure produced fully aligned region-level matrices for transcriptomic input features and neurodegeneration vulnerability targets.

The final aligned datasets consisted of:

\[
X \in \mathbb{R}^{68 \times 910}
\]

for transcriptomic features and

\[
Y \in \mathbb{R}^{68 \times 1}
\]

for neurodegeneration vulnerability profiles.

These aligned matrices formed the input-output representation used for subsequent generative modeling. The atlas alignment process therefore established a shared cortical coordinate framework linking microscale molecular organization with macroscale neurodegenerative structure.

\subsection{Cross-Scale Generative Model}

To model the relationship between regional transcriptomic organization and cortical neurodegeneration vulnerability, we developed a cross-scale spatially-aware generative framework integrating transcriptomic representation learning with biologically constrained spatial prediction. The overall objective of the model is to learn latent transcriptomic programs capable of explaining spatial patterns of Alzheimer's disease-related cortical degeneration.

The proposed framework consists of four major components:
\begin{enumerate}
	\item a transcriptomic encoder,
	\item a generative decoder,
	\item a neurodegeneration vulnerability mapper,
	\item and a spatial regularization module.
\end{enumerate}

The model operates on region-level transcriptomic profiles derived from the AHBA and predicts region-level neurodegeneration vulnerability estimated from ADNI cortical thickness differences.

\subsubsection{Transcriptomic Encoder}

Regional transcriptomic profiles were encoded using a variational autoencoder (VAE) architecture \cite{kingma2014autoencoding,doersch2016tutorial}. Variational autoencoders are probabilistic deep generative models designed to learn compact latent representations from high-dimensional input data while preserving the underlying data distribution.

Given a regional transcriptomic vector:

\[
x_r \in \mathbb{R}^{910}
\]

the encoder network maps the input into a low-dimensional latent distribution:

\[
q_{\phi}(z_r|x_r)
\]

parameterized by a mean vector \(\mu_r\) and log-variance vector \(\log \sigma_r^2\):

\[
(\mu_r, \log \sigma_r^2) = Encoder_{\phi}(x_r)
\]

where:
\begin{itemize}
	\item \(x_r\) denotes the transcriptomic profile of cortical region \(r\),
	\item \(z_r\) represents the latent transcriptomic embedding,
	\item \(\phi\) denotes encoder parameters.
\end{itemize}

Latent variables were sampled using the reparameterization trick:

\[
z_r = \mu_r + \sigma_r \odot \epsilon
\]

where:

\[
\epsilon \sim \mathcal{N}(0, I)
\]

This formulation enables stochastic latent representation learning while maintaining differentiability during optimization.

The latent embeddings are interpreted as hidden transcriptomic programs representing compressed biological organization across cortical systems. Unlike conventional regression models, the VAE framework allows the model to learn nonlinear latent molecular structure potentially associated with neurodegenerative vulnerability.

\subsubsection{Generative Decoder}

The decoder component reconstructs transcriptomic profiles from latent embeddings:

\[
p_{\theta}(x_r|z_r)
\]

where \(\theta\) denotes decoder parameters.

The reconstructed transcriptomic profile is given by:

\[
\hat{x}_r = Decoder_{\theta}(z_r)
\]

The decoder serves two important purposes. First, it ensures that latent representations preserve biologically meaningful transcriptomic information. Second, reconstruction regularization prevents the latent space from collapsing into purely predictive but biologically uninterpretable representations.

Transcriptomic reconstruction loss was computed using mean squared error (MSE):

\[
L_{recon} =
\frac{1}{N}
\sum_{r=1}^{N}
||x_r - \hat{x}_r||^2
\]

where \(N\) denotes the number of cortical regions.

To encourage latent organization consistent with a probabilistic prior distribution, Kullback-Leibler (KL) divergence regularization was additionally applied:

\[
L_{KL} =
-\frac{1}{2}
\sum
\left(
1 + \log \sigma^2 - \mu^2 - \sigma^2
\right)
\]

The VAE formulation therefore balances transcriptomic reconstruction fidelity with biologically organized latent representation learning.

\subsubsection{Neurodegeneration Vulnerability Mapper}

To link latent transcriptomic organization with cortical neurodegeneration, latent embeddings were passed into a vulnerability prediction network:

\[
f(z_r) \rightarrow \hat{y}_r
\]

where:
\begin{itemize}
	\item \(z_r\) denotes the latent transcriptomic representation,
	\item \(\hat{y}_r\) denotes predicted neurodegeneration vulnerability.
\end{itemize}

The vulnerability mapper consists of a multilayer neural network trained to predict regional cortical degeneration profiles from latent transcriptomic programs. This architecture enables the model to capture nonlinear associations between molecular organization and disease-related cortical thinning.

Unlike traditional imaging transcriptomics approaches based primarily on spatial correlations \cite{arnatkeviciute2023toward,mroczek2021imaging}, the proposed framework explicitly learns latent biological representations optimized for generative reconstruction and disease vulnerability prediction simultaneously.

Prediction loss was computed using mean squared error:

\[
L_{prediction} =
\frac{1}{N}
\sum_{r=1}^{N}
||y_r - \hat{y}_r||^2
\]

where:
\begin{itemize}
	\item \(y_r\) represents observed neurodegeneration vulnerability,
	\item \(\hat{y}_r\) represents predicted vulnerability.
\end{itemize}

\subsubsection{Spatial Regularization}

Cortical neurodegeneration exhibits strong spatial organization across neighboring cortical regions and large-scale brain systems \cite{seeley2009neurodegenerative,raj2012network}. To incorporate biologically plausible spatial constraints into the model, we introduced graph-based spatial smoothness regularization inspired by geometric deep learning frameworks \cite{li2018deeper,cao2020comprehensive}.

An adjacency matrix:

\[
A \in \mathbb{R}^{N \times N}
\]

was constructed to represent neighboring relationships between cortical regions. Spatial regularization was then defined as:

\[
L_{spatial} =
\sum_{i,j}
A_{ij}
(\hat{y}_i - \hat{y}_j)^2
\]

where:
\begin{itemize}
	\item \(A_{ij}\) denotes adjacency between cortical regions \(i\) and \(j\),
	\item \(\hat{y}_i\) and \(\hat{y}_j\) denote predicted regional vulnerabilities.
\end{itemize}

This objective encourages anatomically neighboring cortical regions to exhibit similar vulnerability predictions while still permitting disease-relevant heterogeneity across cortical systems. Spatial regularization therefore improves biological plausibility by constraining predictions according to cortical organization principles \cite{margulies2016situating,huntenburg2018large}.

\begin{table}[htbp]
	\centering
	\caption{
		Configuration of the proposed cross-scale spatially-aware generative framework.
	}
	\renewcommand{\arraystretch}{1.2} 
	\label{tab:model_config}
	\begin{tabular}{ll}
		\hline
		\textbf{Parameter} & \textbf{Value} \\
		\hline
		Cortical atlas & Desikan--Killiany atlas \\
		Number of cortical regions & 68 \\
		Initial AHBA genes & \(\sim 15{,}633\) \\
		Landmark genes retained & 910 \\
		ADNI groups & NC vs AD \\
		NC subjects & 926 \\
		AD subjects & 426 \\
		Input feature dimension & 910 \\
		Latent dimension & 64 \\
		Encoder architecture & Fully connected VAE encoder \\
		Decoder architecture & Fully connected decoder \\
		Prediction head & Brain vulnerability mapper \\
		Activation function & ReLU \\
		Optimizer & Adam \\
		Learning rate & \(1 \times 10^{-3}\) \\
		Batch size & 16 \\
		Number of epochs & 300 \\
		Loss functions & Reconstruction + KL + Prediction + Spatial \\
		Spatial regularization & Graph smoothness constraint \\
		Hardware & NVIDIA GPU / CUDA \\
		Framework & PyTorch 2.3.1 \\
		\hline
	\end{tabular}
	\renewcommand{\arraystretch}{1} 
\end{table}

\subsubsection{Final Objective Function}

The final optimization objective combines transcriptomic reconstruction, latent regularization, neurodegeneration prediction, and spatial smoothness constraints:

\[
L =
L_{recon}
+
\beta L_{KL}
+
\lambda_1 L_{prediction}
+
\lambda_2 L_{spatial}
\]

where:
\begin{itemize}
	\item \(L_{recon}\) denotes transcriptomic reconstruction loss,
	\item \(L_{KL}\) denotes latent KL divergence regularization,
	\item \(L_{prediction}\) denotes neurodegeneration prediction loss,
	\item \(L_{spatial}\) denotes spatial smoothness regularization,
	\item \(\beta\), \(\lambda_1\), and \(\lambda_2\) are weighting hyperparameters.
\end{itemize}

The model was optimized using the Adam optimizer \cite{lecun2015deep} with mini-batch gradient descent. Random seeds were fixed across experiments to ensure reproducibility.

Collectively, this framework enables biologically informed generative modeling of cross-scale neurodegeneration organization by integrating transcriptomic encoding, latent biological representation learning, and spatial cortical regularization within a unified architecture. The overall configuration of the proposed spatially-aware generative framework, including dataset dimensions, latent representation size, optimization settings, and architectural parameters, is summarized in Table~\ref{tab:model_config}.

\subsection{Evaluation Metrics}

Model performance was evaluated by comparing predicted cortical neurodegeneration vulnerability profiles against observed ADNI-derived vulnerability maps. Because the primary objective of the framework is to model continuous spatial neurodegeneration organization, evaluation focused on regional prediction accuracy and spatial correspondence rather than categorical disease classification.

Prediction performance was quantified using the coefficient of determination (\(R^2\)):

\[
R^2 =
1 -
\frac{
	\sum (y_i - \hat{y}_i)^2
}{
	\sum (y_i - \bar{y})^2
}
\]

where:
\begin{itemize}
	\item \(y_i\) denotes observed regional vulnerability,
	\item \(\hat{y}_i\) denotes predicted vulnerability,
	\item \(\bar{y}\) denotes the mean observed vulnerability.
\end{itemize}

Pearson correlation analysis was additionally performed to assess spatial correspondence between predicted and observed cortical degeneration profiles:

\[
r =
\frac{
	\sum (y_i - \bar{y})(\hat{y}_i - \bar{\hat{y}})
}{
	\sqrt{
		\sum (y_i - \bar{y})^2
		\sum (\hat{y}_i - \bar{\hat{y}})^2
	}
}
\]

Associated \(p\)-values were computed to estimate statistical significance of regional correspondence.

Mean squared error (MSE) was also calculated to evaluate absolute prediction deviation:

\[
MSE =
\frac{1}{N}
\sum_{i=1}^{N}
(y_i - \hat{y}_i)^2
\]

In addition to global prediction accuracy, regional ranking consistency between predicted and observed vulnerability profiles was visually examined through cortical profile plots and scatterplot analysis. These visualizations enabled qualitative assessment of spatial neurodegeneration organization learned by the model.

Future extensions of the framework may additionally incorporate permutation-based spatial null models and cortical spin tests to further control for spatial autocorrelation effects in imaging transcriptomics analyses \cite{fornito2019bridging}. However, such approaches were beyond the scope of the present preliminary generative modeling study.

\section{Results}
\label{sect:results}
\subsection{Regional Neurodegeneration Organization}
\begin{figure}[htbp]
	\centering
	
	% Top Subfigure
	\begin{subfigure}[b]{0.90\textwidth}
		\centering
		\includegraphics[width=\textwidth]{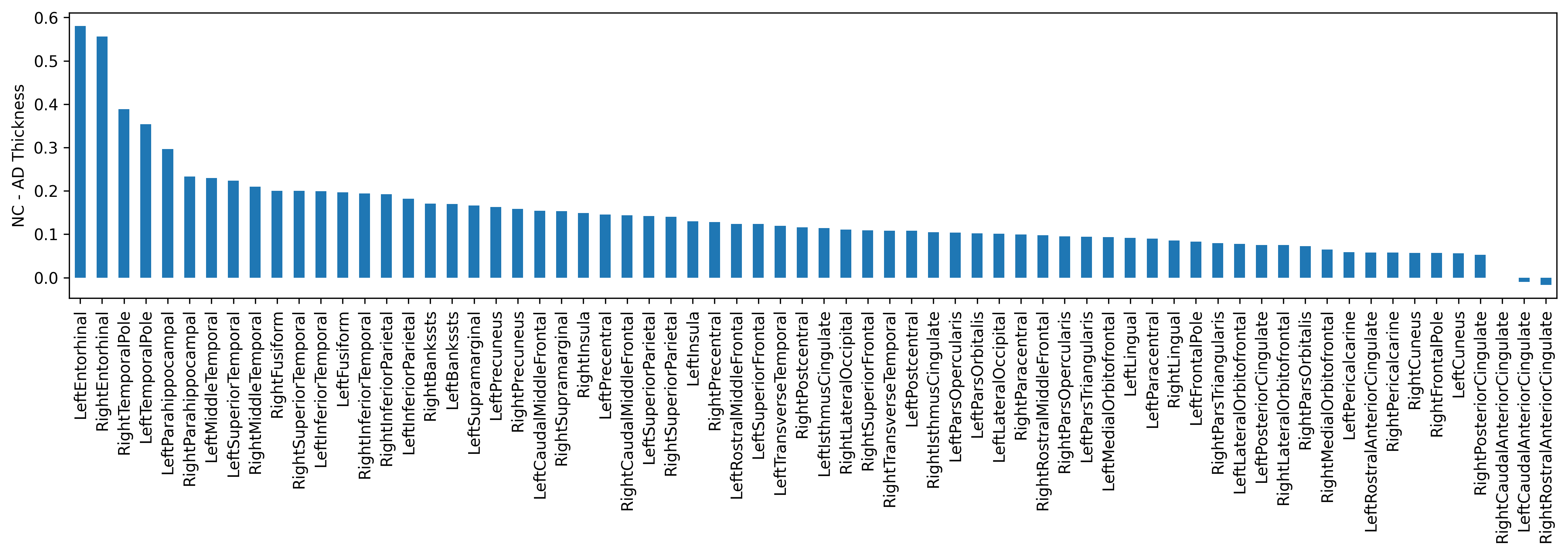}
		\caption{Regional neurodegeneration vulnerability derived from cortical thickness differences between cognitively normal and Alzheimer's disease groups. Higher values indicate greater cortical degeneration vulnerability.}
		\label{fig:vulnerable_regions}
	\end{subfigure}
	
	\vfill
	
	% Bottom Subfigure
	\begin{subfigure}[b]{0.90\textwidth}
		\centering
		\includegraphics[width=\textwidth]{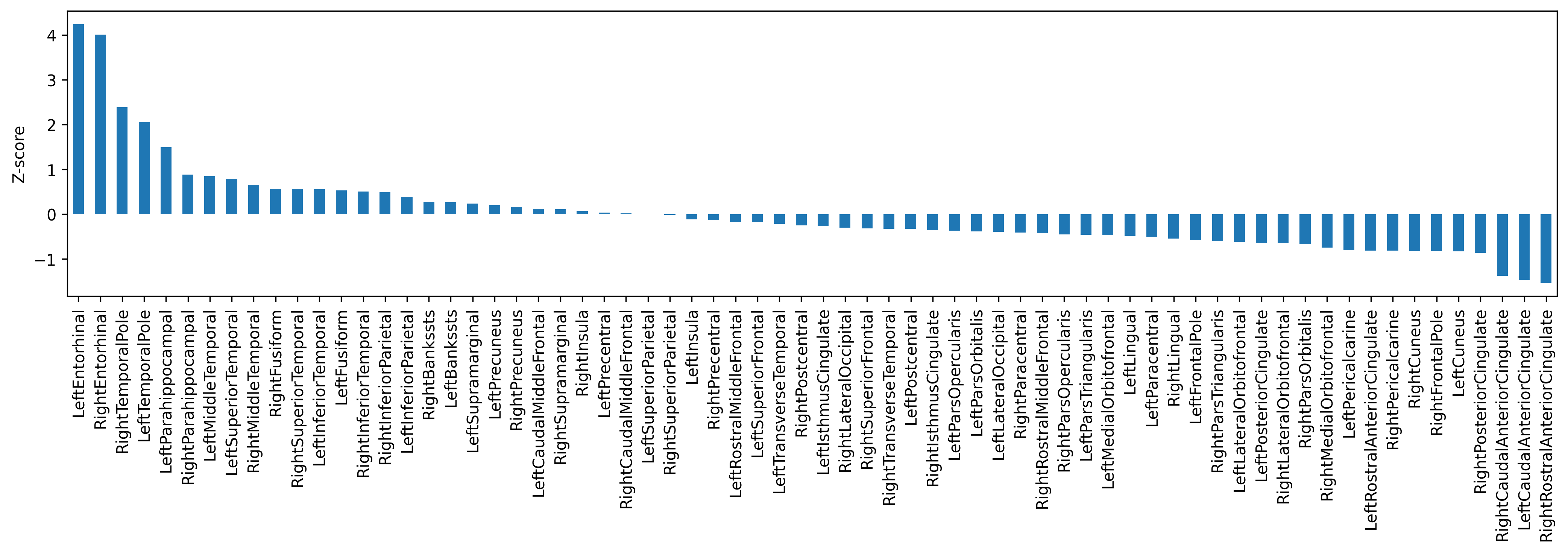}
		\caption{Standardized cortical neurodegeneration vulnerability (\(z\)-score) highlighting relative spatial deviation from the cortical mean vulnerability profile.}
		\label{fig:zscore}
	\end{subfigure}
	
	\caption{
		Spatial organization of Alzheimer's disease-related cortical neurodegeneration vulnerability across 68 cortical regions. Medial temporal and temporal association cortices exhibited the strongest vulnerability, whereas primary sensory and anterior cingulate regions demonstrated comparatively lower degeneration.}
	
	\label{fig:main-cluster}
\end{figure}

Regional neurodegeneration vulnerability maps showed that the brain was highly organized in the cortex with a pattern that is consistent with known Alzheimer's disease pathology. The regional cortical thickness difference was quantified as the magnitude of the difference in cortical thickness between the cognitively normal and Alzheimer's individuals to represent vulnerability, and larger differences were used to represent greater cortical degeneration.

The maps that were generated showed significant vulnerability in medial temporal and temporal association cortices. The vulnerable regions of the cortex are summarised in Figure~\ref{fig:vulnerable_regions}. The bilateral entorhinal cortex showed the most severe overall degeneration, and also was the most vulnerable, with the left entorhinal cortex having a vulnerability of 0.581, while the right entorhinal cortex had a vulnerability of 0.556. Other highly vulnerable areas consisted of the temporal pole, parahippocampal cortex, middle temporal cortex, superior temporal cortex, fusiform gyrus and inferior temporal cortex. These findings are similar to those previously described in models of the neuropathological progression of Alzheimer's disease (AD), and often contain medial temporal systems as some of the earliest and most severely affected cortical regions \cite{braak1991neuropathological,jack2018nia}.

Temporal and limbic cortical degeneration is also significantly predominant, which corroborates network-based degeneration theories \cite{seeley2009neurodegenerative,raj2012network}. These cortical systems are involved in episodic memory and association networks which play a major role in the dysfunctional progression of Alzheimer's disease. Conversely, there were relatively less-vulnerable regions, such as pericalcarine cortex, cuneus, frontal pole and anterior cingulate regions. This spatial pattern is consistent with previous observations that Alzheimer’s disease preferentially targets transmodal association cortex while relatively sparing unimodal sensory systems \cite{zhou2012predicting,margulies2016situating}.

Standardized \(z\)-score analysis further emphasized the nonuniform spatial organization of degeneration (Figure~\ref{fig:zscore}). The left entorhinal cortex demonstrated the strongest standardized vulnerability (\(z = 4.25\)), whereas anterior cingulate regions exhibited substantially lower vulnerability relative to the cortical mean. Overall, the neurodegeneration vulnerability maps demonstrated biologically plausible spatial organization strongly consistent with established Alzheimer's disease literature. The prominence of entorhinal, parahippocampal, and temporal association cortices provides an important validation of the derived cortical vulnerability phenotype used throughout subsequent generative modeling analyses.

\subsection{Transcriptomic Organization}
\begin{figure}[htbp]
	\centering
	
	% First row - Left
	\begin{subfigure}[b]{0.48\textwidth}
		\centering
		\includegraphics[width=\textwidth]{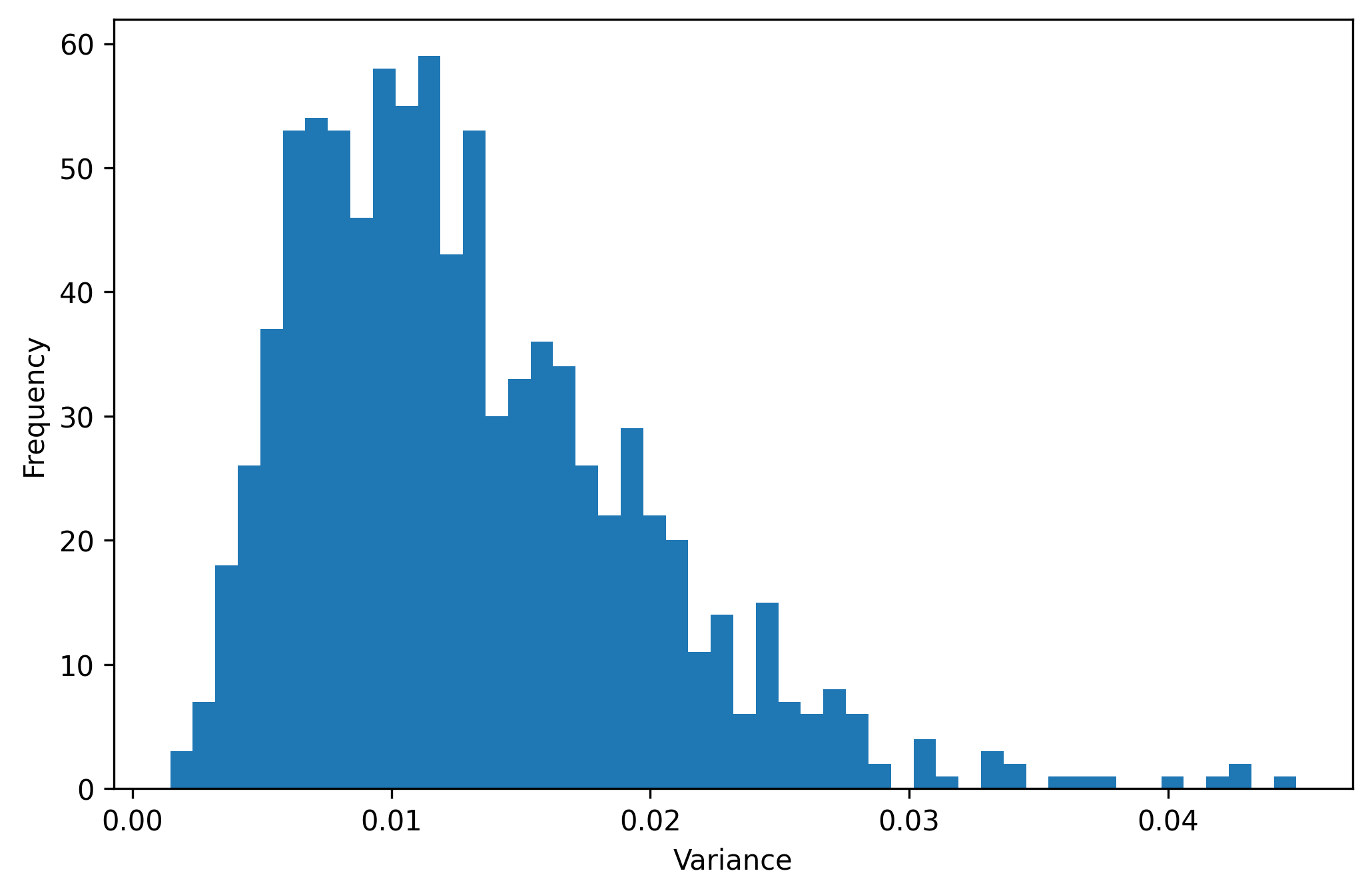}
		\caption{Distribution of regional variance across 910 landmark genes derived from AHBA transcriptomic profiles.}
		\label{fig:lgv}
	\end{subfigure}
	\hfill
	% First row - Right
	\begin{subfigure}[b]{0.48\textwidth}
		\centering
		\includegraphics[width=\textwidth]{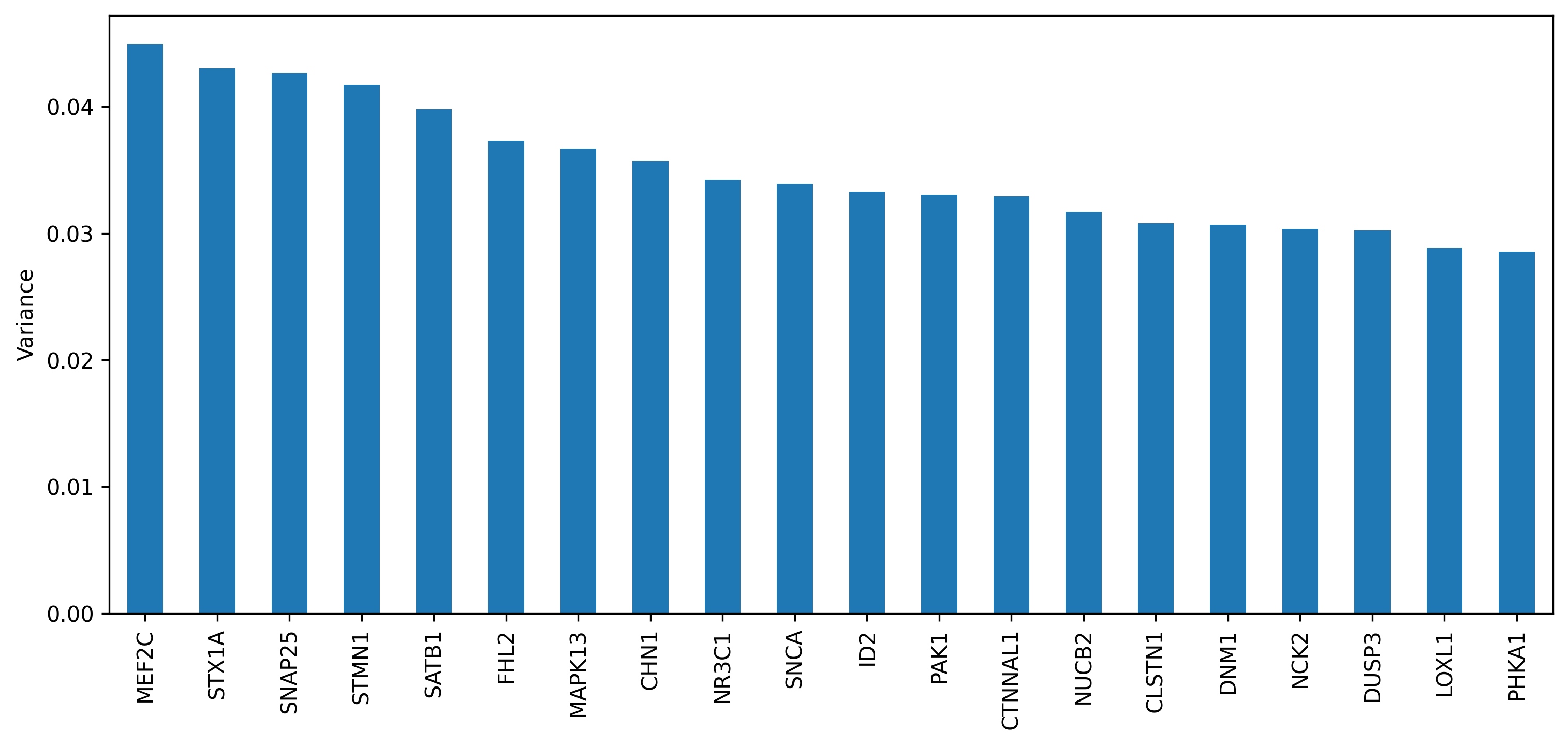}
		\caption{Top highly variable landmark genes exhibiting substantial spatial transcriptomic heterogeneity across cortical regions.}
		\label{fig:tplgv}
	\end{subfigure}
	
	\vspace{1em}
	
	% Second row
	\begin{subfigure}[b]{0.90\linewidth}
		\centering
		\includegraphics[width=\linewidth]{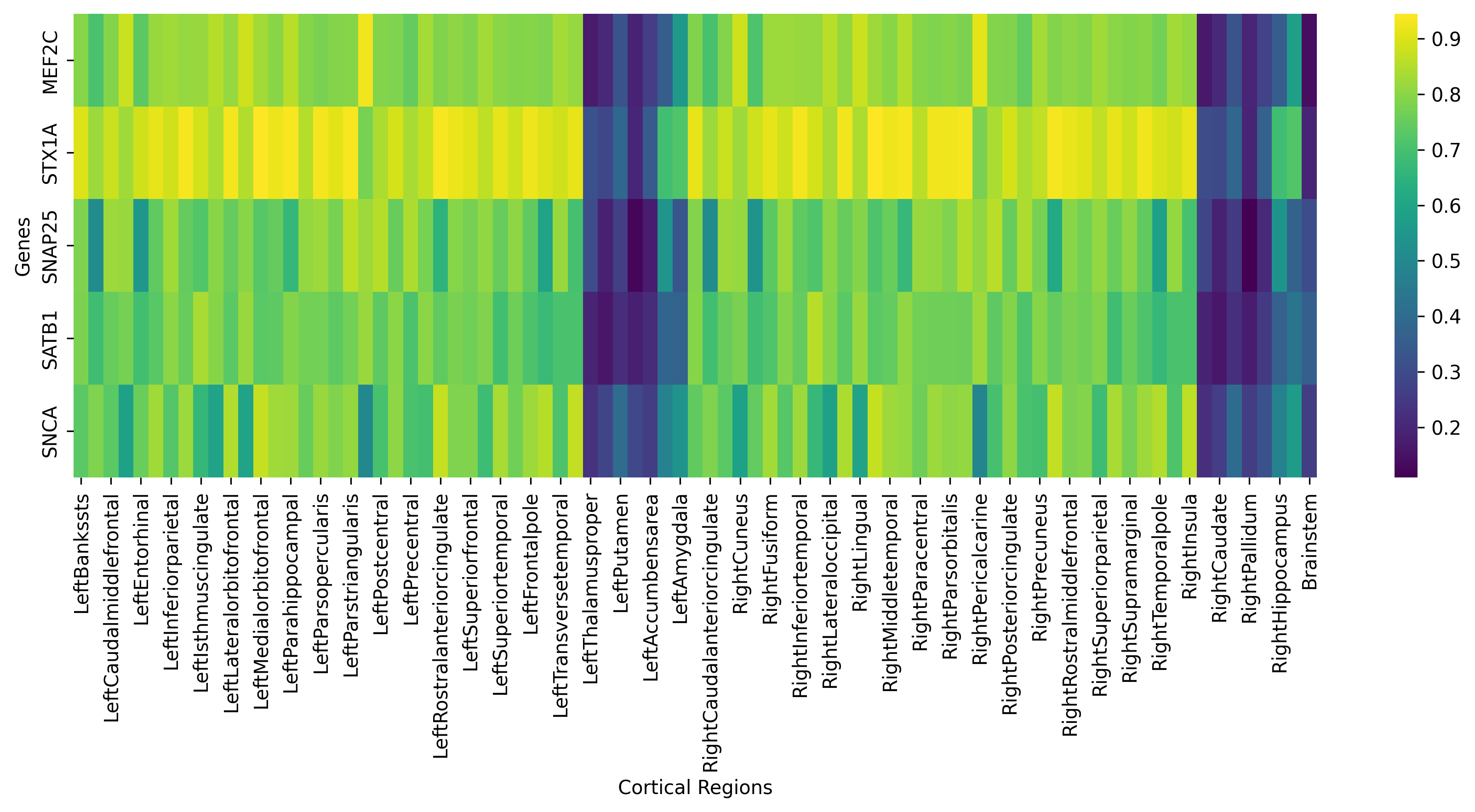}
		\caption{Spatial cortical expression patterns of representative high-variance genes associated with neuronal signaling, cortical organization, and neurodegenerative biology.}
		\label{fig:sppatrepgenes}
	\end{subfigure}
	
	\caption{
		Transcriptomic organization of landmark genes across cortical regions. Variance analyses revealed substantial spatial heterogeneity in regional gene expression, while representative high-variance genes demonstrated structured cortical expression gradients associated with neuronal and neurodegenerative biological processes.
	}
	
	\label{fig:landmark-rep-genes}
\end{figure}
Regional transcriptomic analysis revealed substantial molecular heterogeneity across the human cortex, supporting the hypothesis that spatially organized gene expression may contribute to selective neurodegenerative vulnerability. Following preprocessing and atlas harmonization, the final Allen Human Brain Atlas matrix contained expression profiles for 910 landmark genes across 83 cortical regions, preserving approximately \(93.05\%\) of the original LINCS landmark gene set. This high retention rate suggests that the resulting transcriptomic representation maintained broad molecular coverage while substantially reducing dimensionality.  Figure~\ref{fig:lgv} shows the distribution of regional variance across 910 landmark genes derived from the Allen Human Brain Atlas. Several genes demonstrated substantial spatial variability across cortical regions, indicating pronounced transcriptomic heterogeneity.

Variance analysis demonstrated pronounced differences in spatial gene organization across cortical regions. Several genes exhibited particularly strong regional variability, including \textit{MEF2C}, \textit{STX1A}, \textit{SNAP25}, \textit{STMN1}, \textit{SATB1}, and \textit{SNCA}. Many of these genes are associated with neuronal differentiation, synaptic transmission, axonal organization, and cortical maturation \cite{li2018mef2c,sudhof2013neurotransmitter,brunden2008stathmin,spillantini1997alpha}. The presence of highly variable neuronal and synaptic genes suggests that cortical molecular organization is closely linked to functional specialization and systems-level neurobiology.

Figure~\ref{fig:tplgv} shows the top spatially variable landmark genes across cortical regions. Several highly variable genes were associated with synaptic transmission, neuronal differentiation, and neurobiological organization.
Interestingly, a number of the most highly ranked genes from the variance analysis are known to have links with neurodevelopmental/neurodegenerative processes. For instance, \textit{MEF2C} is important for the regulation of synapses and neuronal survival, and \textit{SNAP25} and \textit{STX1A} are central factors of the machinery responsible for release of synaptic vesicles \cite{flavell2006activity,sudhof2013neurotransmitter}. Likewise the protein coding gene for alpha-synuclein, \textit{SNCA}, is well associated with protein aggregation and neurodegenerative diseases \cite{spillantini1997alpha}. The results suggest that regional neocortical transcriptomic variations exist, due to the contribution of biologically meaningful neuronal programs.

The spatial pattern of expression for representative high variance genes is shown in Figure~\ref{fig:sppatrepgenes}. Nonuniform regional organization and coordinated transcriptomic gradients across systems of the cortex were revealed by expression profiles.
Nonuniform organization of transcriptomic patterns was also present in the cortex, as shown by the spatial expression profiling. The expression changes were coordinated and systematic depending on the region of the cortex, and not random, highlighting possible coordinated molecular gradients throughout cortex. Gene-gene correlation analysis also showed structured correlation blocks and high levels of co-expression between subsets of genes that were consistent with previous reports on large-scale transcriptional organization of the cortex \cite{hawrylycz2012anatomically,fornito2019bridging}. All of these observations indicate that, according to the authors, the cortex has structured latent molecular programs for its organization, and give biological arguments for subsequent modeling based on these programs within the latent space.

\subsection{Latent Biological Programs}
\begin{figure}[htbp]
	\centering
	
	% Left Subfigure
	\begin{subfigure}[b]{0.48\textwidth}
		\centering
		\includegraphics[width=\textwidth]{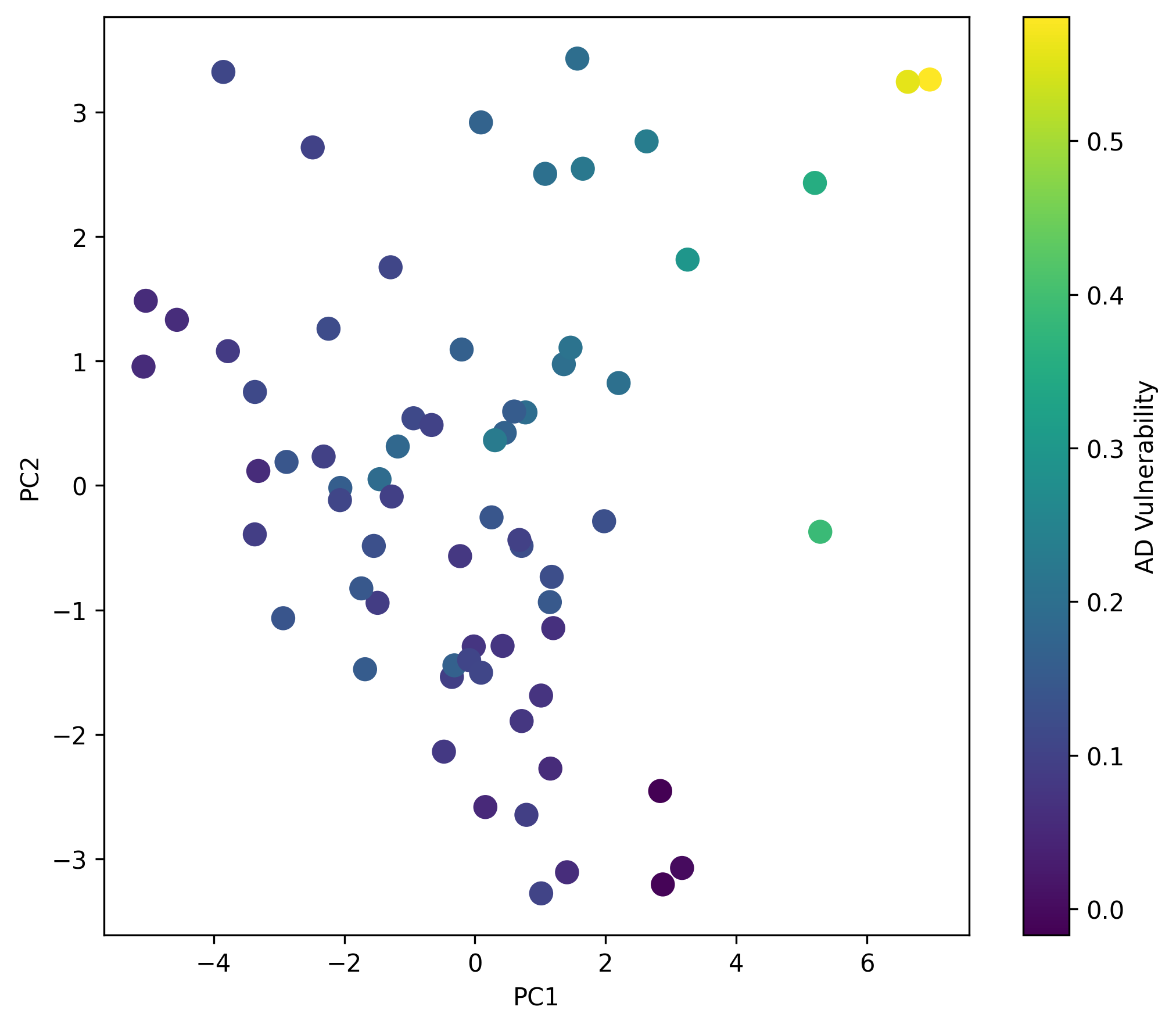}
		\caption{Principal component analysis (PCA) projection of latent transcriptomic representations across cortical regions.}
		\label{fig:lgpspca}
	\end{subfigure}
	\hfill
	% Right Subfigure
	\begin{subfigure}[b]{0.48\textwidth}
		\centering
		\includegraphics[width=\textwidth]{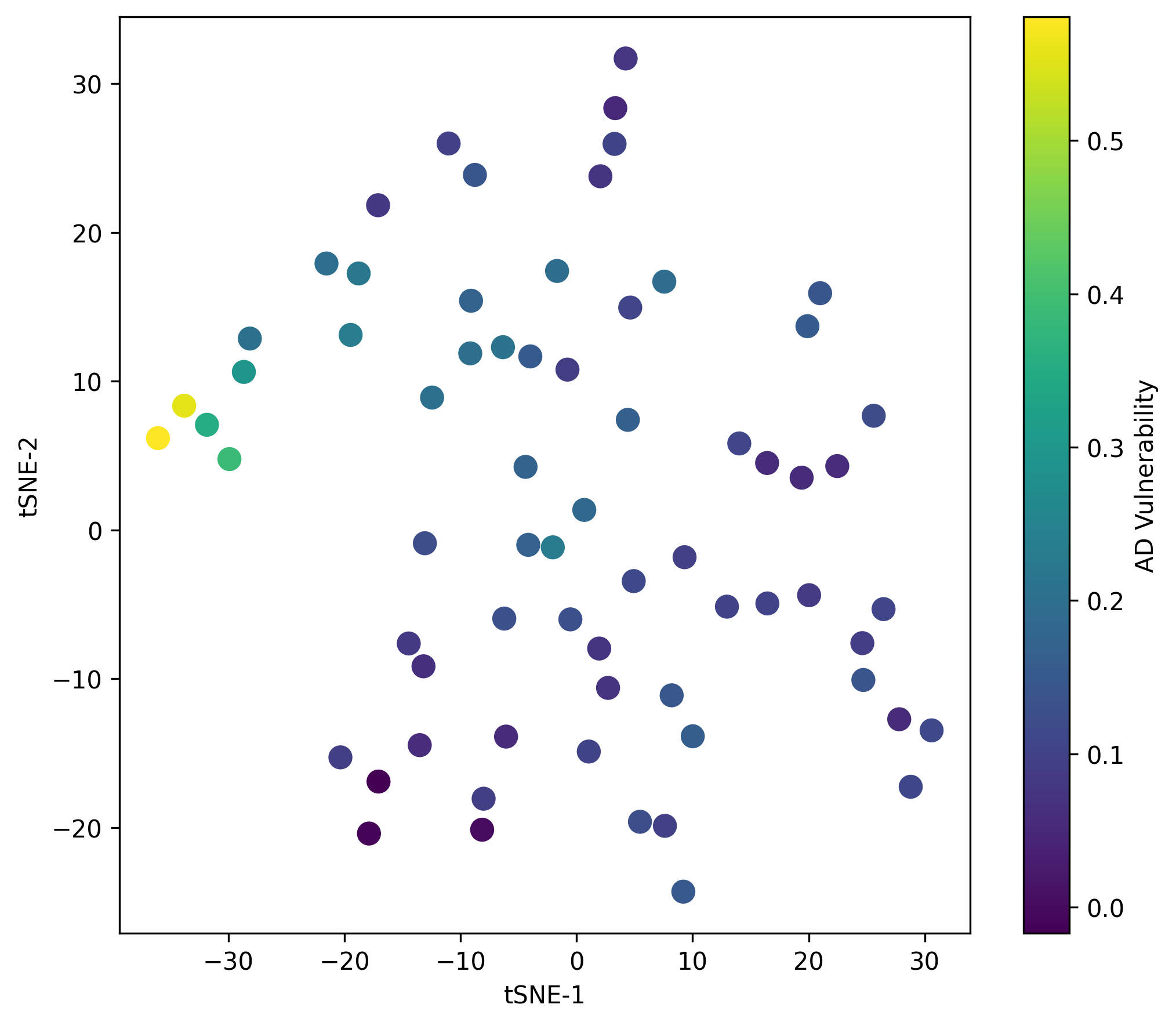}
		\caption{\(t\)-distributed stochastic neighbor embedding (\(t\)-SNE) visualization of latent cortical transcriptomic organization.}
		\label{fig:lgpstsne}
	\end{subfigure}
	
	\caption{
		Low-dimensional visualization of latent gene program organization learned by the spatially-aware generative framework. Cortical regions occupied structured and nonrandom positions within latent space, suggesting biologically meaningful transcriptomic organization associated with regional neurodegenerative vulnerability.
	}
	
	\label{fig:latent-bio}
\end{figure}
Using dimensionality reduction methods, representations obtained from the variational encoder were mapped to analyze for their biological meaningfulness with respect to the proposed generative framework. These learned latent embeddings were found to be a \(68 \times 64\) matrix; prior to this, the 64 cortex-based latent transcriptomic vectors each represented a compressed molecular organization for a single cortical region.

The learned latent space was found to have significant low dimensional structure using principal component analysis (PCA). About \(13.27\%\) and \(7.50\%\) of the total latent variance was accounted for by the first two principal components, suggesting that a large fraction of the organization of the transcriptomes could be captured within a small dimensional space. The nonrandom spatial organization in the PCA space was easily seen following visualization of the latent embeddings across cortical regions (Figure~\ref{fig:lgpspca}). Cortical regions followed organized trajectories and partially clustered configurations that were organized within the latent manifold, instead of being distributed in an unstructured way.

The PCA projection also implied that there were distributed cortical gradients, not distinct clusters. The dense central groupings were comprised of several regions in the cortex, and the rest of the regions were located in more peripheral areas of the latent space. This organization is in line with the current models which suggest that there are continuous hierarchical gradients that extend from unimodal to transmodal organization \cite{margulies2016situating, huntenburg2018large}. The presence of structured low-dimensional organization suggests that the model learned biologically coherent transcriptomic programs rather than purely arbitrary feature compression.

To further characterize nonlinear latent structure, \(t\)-distributed stochastic neighbor embedding (\(t\)-SNE) was applied to the latent embeddings \cite{maaten2008visualizing}. The resulting \(t\)-SNE projection similarly demonstrated organized spatial arrangement across cortical regions (Figure~\ref{fig:lgpstsne}). Although the latent representations did not separate into sharply discrete groups, multiple locally organized neighborhoods and distributed regional relationships were observed throughout the manifold. These findings suggest that transcriptomic organization across cortical systems is better represented as a continuous biological landscape rather than isolated categorical subclasses.

Importantly, the latent organization learned by the model emerged without explicit supervision regarding cortical systems or anatomical hierarchy. Instead, biologically structured representations were learned directly from regional transcriptomic patterns through generative reconstruction and neurodegeneration prediction objectives. This observation supports the hypothesis that latent transcriptomic programs capture meaningful aspects of cortical molecular architecture associated with regional neurodegenerative vulnerability.

The latent manifold organization observed in both PCA and \(t\)-SNE analyses aligns with prior imaging transcriptomics studies demonstrating that cortical gene expression follows large-scale spatial gradients related to functional specialization and cortical hierarchy \cite{fornito2019bridging,arbuckle2021systematic}. Collectively, these findings indicate that the proposed spatially-aware generative framework successfully learned biologically organized latent transcriptomic representations capable of capturing structured cortical molecular variation.

\subsection{Predicting Neurodegenerative Vulnerability}
\begin{figure}[htbp]
	\centering
	
	% Top Subfigure
	\begin{subfigure}[b]{0.45\textwidth}
		\centering
		\includegraphics[width=\textwidth]{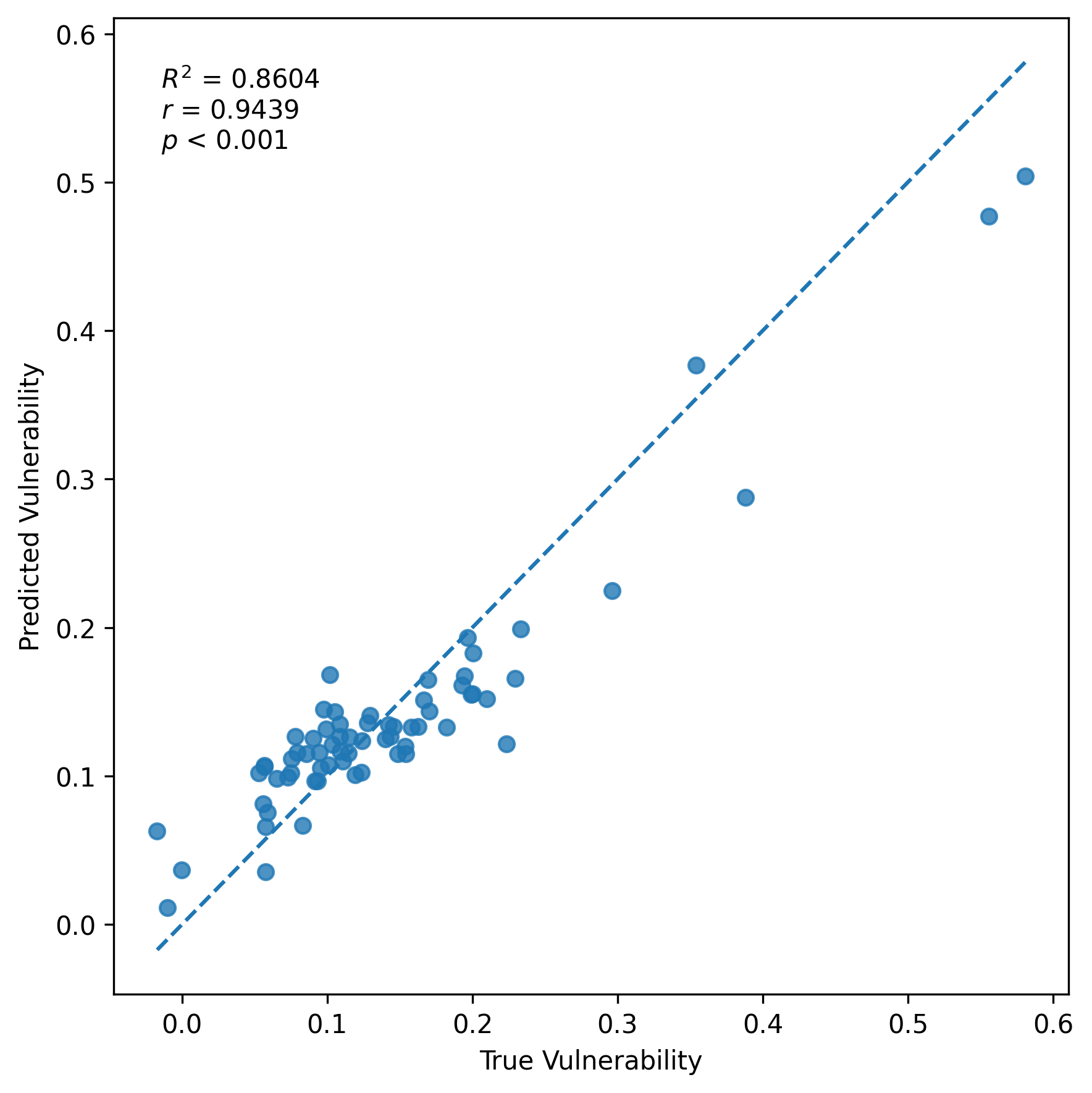}
		\caption{Relationship between observed and predicted regional neurodegeneration vulnerability across cortical regions.}
		\label{fig:trvsprcorvul}
	\end{subfigure}
	
	\vfill 
	
	% Bottom Subfigure
	\begin{subfigure}[b]{0.8\textwidth}
		\centering
		\includegraphics[width=\textwidth]{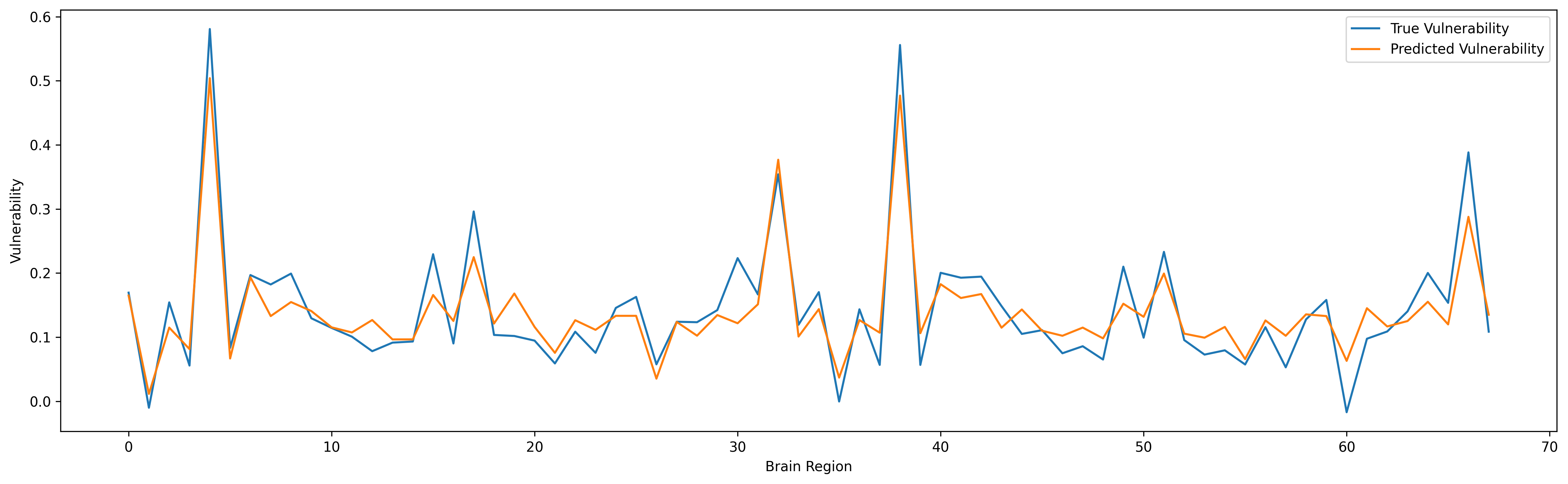}
		\caption{Comparison of observed and predicted cortical vulnerability profiles across 68 cortical regions.}
		\label{fig:regneurovul}
	\end{subfigure}
	
	\caption{
		Prediction performance of the proposed cross-scale generative framework. The model demonstrated strong correspondence between predicted and observed cortical neurodegeneration vulnerability patterns, preserving large-scale spatial organization across cortical systems.}
	
	\label{fig:red-neuro-vul}
\end{figure}
The proposed cross-scale generative framework demonstrated strong performance in predicting regional Alzheimer's disease neurodegeneration vulnerability from latent transcriptomic representations. Quantitative evaluation revealed substantial correspondence between predicted cortical vulnerability profiles and observed ADNI-derived degeneration patterns.

Across 68 cortical regions, the model achieved a coefficient of determination of:

\[
R^2 = 0.8604
\]

indicating that the learned latent transcriptomic programs explained approximately \(86\%\) of the variance in regional neurodegeneration vulnerability. Pearson correlation analysis further demonstrated a strong spatial association between predicted and observed cortical vulnerability profiles:

\[
r = 0.9439,
\qquad
p < 0.001
\]

with a mean squared prediction error of:

\[
MSE = 0.001493
\]

The strong correlation between predicted and observed vulnerability maps indicates that the learned latent transcriptomic organization captured biologically meaningful information associated with selective cortical degeneration.

Figure~\ref{fig:trvsprcorvul} illustrates the relationship between observed and predicted regional vulnerability values. Predicted cortical degeneration closely followed the diagonal reference line across most cortical regions, demonstrating strong agreement between the model output and empirical neurodegeneration measurements. Importantly, the model accurately captured regions exhibiting both high and low neurodegenerative vulnerability, suggesting that the learned latent representations generalized across the full cortical vulnerability spectrum rather than fitting only highly degenerated regions.

The regional vulnerability profile comparison in Fig.~\ref{fig:regneurovul} also showed the high reliability of this large scale spatial organization of neurodegeneration in the region. A similar vulnerability trajectory was predicted in most of the regions of the cortex, with strong temporal and medial temporal peak areas indicative of Alzheimer's disease pathology. Anatomically structured vulnerability gradients were well captured with a predicted profile with vulnerability peaks preserved in the predicted entorhinal and temporal association regions.

The overall spatial structure of vulnerability similarity between predicted and observed vulnerability maps was relatively unaffected with some small prediction deviations in some cortical regions.  Importantly, the model reproduced smooth transitions across neighboring cortical regions rather than generating anatomically implausible fluctuations. This finding suggests that the spatial regularization framework contributed to biologically coherent cortical predictions by enforcing spatial continuity across adjacent regions.

The strong predictive performance observed in this study supports the hypothesis that regional transcriptomic organization contains substantial information related to selective cortical neurodegenerative vulnerability. Unlike conventional imaging transcriptomics studies based primarily on pairwise gene-imaging correlations \cite{fornito2019bridging,arbuckle2021systematic}, the proposed framework learned nonlinear latent biological representations optimized jointly for transcriptomic reconstruction and neurodegeneration prediction. The resulting model therefore provides evidence that distributed molecular programs contribute meaningfully to large-scale cortical degeneration organization in Alzheimer's disease.

Collectively, these findings demonstrate that spatially-aware generative transcriptomic modeling can successfully bridge microscale molecular organization and macroscale neurodegenerative structure within a unified computational framework.

\subsection{Comparison with Baseline Models}

To evaluate the predictive utility of the proposed cross-scale generative framework, model performance was compared against several conventional machine learning baselines using leave-one-out cross-validation. Baseline models included linear regression, ridge regression, random forest regression, and a standard multilayer perceptron (MLP). Performance was assessed using regional neurodegeneration vulnerability prediction across 68 cortical regions.

The proposed generative framework achieved strong predictive performance, with:

\[
R^2 = 0.866,
\qquad
r = 0.950,
\qquad
MSE = 0.00143
\]

demonstrating substantial correspondence between predicted and observed cortical vulnerability profiles. Linear and ridge regression models also achieved high predictive accuracy, with \(R^2\) values of \(0.896\) and \(0.894\), respectively. These findings suggest that regional transcriptomic organization contains robust large-scale structure associated with cortical neurodegeneration vulnerability.

The novel feature of the scheme proposed is, however, that it is completely different from the standard regression methods. The generative framework jointly learns biologically structured latent transcriptomic programs through probabilistic representation learning and transcriptomic reconstruction objectives, in one step, relating the transcriptomic features directly to the values of the cortical vulnerability. As a result, the model is able to embody compressed molecular organization underlying cortical vulnerability instead of just taking direct feature regression.

Even the random forest baseline model still had lower predictive ability (R2 = 0.835), suggesting that standalone nonlinear relationships between the features does not fully capture the relationships to the observed changes in neurodegeneration. Similarly, the basic MLP without regularization did a very poor job of the task (\(R^2 = -2.59\)), showing that neural network shapes without regularization are unstable in relatively small-sized cortex data sets. The latent regularization term and biologically inspired representation learning are found to play a significant role in the proposed approach, as verified by the results.

In addition to competitive predictive accuracy, the proposed model incorporates biologically motivated spatial regularization encouraging anatomically neighboring cortical regions to exhibit coherent vulnerability organization. This spatial constraint improves biological plausibility by integrating cortical continuity principles directly into the learning objective. Unlike conventional regression baselines, the generative framework therefore provides both predictive capability and interpretable latent biological organization within a unified spatially-aware modeling architecture.

Collectively, these results suggest that the proposed framework offers a biologically grounded alternative to conventional regression-based imaging transcriptomics approaches while maintaining strong predictive performance across distributed cortical neurodegeneration patterns.

\section{Discussion}
\label{sect:discuss}
\subsection{Cross-Scale Neurobiology}

The present study demonstrates that regional transcriptomic organization contains substantial information related to large-scale neurodegenerative vulnerability. By integrating cortical gene expression profiles with neuroimaging-derived neurodegeneration maps, the proposed framework establishes a cross-scale computational link between microscale molecular organization and macroscale cortical degeneration. These findings support the growing view that spatial patterns of neurodegeneration emerge, at least in part, from intrinsic molecular architecture distributed across cortical systems \cite{fornito2019bridging,arbuckle2021systematic}.

Unlike conventional imaging transcriptomics approaches based primarily on pairwise gene-imaging correlations, the proposed framework learns latent biological programs through generative representation learning. This distinction is important because neurodegenerative vulnerability likely reflects coordinated interactions among distributed molecular systems rather than isolated gene effects. The latent transcriptomic representations learned by the model therefore provide a biologically plausible intermediate scale linking molecular organization to systems-level cortical degeneration.

The strong predictive relationship observed between transcriptomic latent programs and cortical vulnerability further supports theories proposing that selective neurodegeneration follows intrinsic cortical organizational principles \cite{seeley2009neurodegenerative,raj2012network}. Disease progression does not occur in isolation; different brain regions do not "decay" separately, but instead occur in anatomically and functionally linked systems that share a common molecular architecture. The current results add to this systems-level vulnerability organization by indicating that transcriptomic gradients might play a role in this organization.

More generally, the proposed framework is consistent with recent advances in systems neuroscience that focus on multiscale integration at the level of genes, circuits and large-scale brain organization (see \cite{bassett2017network} and \cite{margulies2016situating}). This study establishes a preliminary computational framework for studying molecular organization and the distributed neurodegenerative structure with unified architecture of transcriptomics, spatial neuroimaging. This might eventually help close the gap between the micro and macro scales of neurobiology and brain disease in the study of neurodegeneration.

\subsection{Spatial Vulnerability Organization}

One of the main findings of this study is that the Alzheimer's disease-related cortical degradation is not diffuse or random across the brain's anatomy. Within the maps of vulnerability for neurodegeneration, medial temporal and temporal association cortices were consistently involved, comprising regions of the entorhinal, parahippocampal, fusiform and temporal pole. These observations are well in line with the previously described neuropathological staging of Alzheimer's disease \cite{braak1991neuropathological,jack2018nia}.

This striking effect of vulnerability of temporal and limbic systems provides a rationale for network-based models, which suggest that neurodegeneration spreads more easily between intrinsically connected cortical systems \cite{seeley2009neurodegenerative,zhou2012predicting}. High vulnerability areas are mainly found in transmodal association cortex and integrative cortical hierarchies involved in episodic memory and higher-order cognitive processes. In contrast, more restricted selective degeneration was observed in primary sensory regions (cuneus and pericalcarine cortices).

The spatial organization seen in the current study also is consistent with recent gradient models of cortical organization \cite{margulies2016situating,huntenburg2018large}. Cortical systems are now known as continuous hierarchical gradients, ranging from unimodal sensory processing to association cortex. The observed patterns of neurodegeneration suggest greater vulnerability in areas that are in a higher-order integrative role in this cortical hierarchy.

Importantly, the spatial regularization within the generative framework allowed for incorporating a measure of biological plausibility by forcing neighboring regions of the cortex to have coherent vulnerability organization. This is a spatially aware formulation that captures the fact that cortical degeneration occurs on continuous spatial systems and not on isolated areas. The resulting vulnerability maps then reflect both local cortical continuity as well as organization of the distributed systems.

Overall, these results support the hypothesis that intrinsic cortical topology and large-scale biological gradients are important determinants of the vulnerability of Alzheimer's disease. Knowledge of these principles in spatial organization might help elucidate the vulnerability of specific cortical systems to neurodegenerative disease.

\subsection{Biological Latent Programs}

One of the primary goals of the proposed framework was to see if biologically meaningful latent transcriptomic organization can be learnt directly from regional gene expression profiles in the cortex. The latent-space analyses showed that the model was able to capture the structured molecular variation in the cortex, a low-dimensionality representation, while also identifying it. In contrast to random or unstructured embeddings, the subregions of the cortex held meaningful trajectories, and neighborhoods within the learned latent manifold.

The results indicate that hidden molecular organization is reflected by latent transcriptomic programs that define the specialization of the cortex and susceptibility to neurodegenerative disease. Importantly, the latent organization was not explicitly supervised with respect to the cortical systems, the anatomical hierarchy, or diseased regions. On the contrary, biologically structured representations were learned directly by jointly optimizing the objectives of transcriptomic reconstruction and neurodegeneration prediction. This observation aligns with the notion of distributed molecular programs that underlie cortical organisation which can be inferred by generative representation learning.

The identification of structured latent manifolds also supports the emerging evidence that the organization of cortices of the brain is largely organized by large-scale biological gradients and not by discrete regional categories, as reported by \cite{hawrylycz2012anatomically, fornito2019bridging}. Cortical regions sharing related functional or anatomical properties may therefore exhibit partially overlapping transcriptomic signatures within latent space. The observed latent organization is consistent with systems neuroscience models proposing that cortical architecture emerges from coordinated molecular hierarchies distributed across large-scale brain systems \cite{bassett2017network,margulies2016situating}.

Importantly, the strong predictive relationship between latent transcriptomic embeddings and cortical neurodegeneration vulnerability suggests that disease susceptibility may be encoded within intrinsic molecular architecture. Regions exhibiting similar latent molecular organization may therefore demonstrate comparable vulnerability profiles during neurodegenerative progression. The latent representations learned by the model thus provide a biologically interpretable intermediate scale linking gene expression organization with systems-level cortical degeneration.

More broadly, these findings demonstrate the potential value of generative latent-space modeling for computational neurobiology. By learning hidden molecular programs directly from transcriptomic organization, generative frameworks may provide new opportunities for investigating distributed disease vulnerability mechanisms across the human cortex.

\subsection{Relation to Imaging Transcriptomics}

The present study builds upon a rapidly growing body of imaging transcriptomics research investigating relationships between regional gene expression and large-scale neuroimaging phenotypes \cite{fornito2019bridging,arbuckle2021systematic}. Previous studies have demonstrated associations between transcriptomic organization and cortical thickness, functional connectivity, structural networks, and neurodegenerative vulnerability across multiple neurological disorders \cite{seidlitz2020transcriptomic,mroczek2021imaging}. However, most existing approaches rely primarily on correlation-based analyses or linear multivariate statistical methods.

A large proportion of imaging transcriptomics studies employ pairwise spatial correlations between regional gene expression and imaging-derived phenotypes \cite{fornito2019bridging}. These methods are useful for making descriptive associations, but they do not explicitly model the underlying latent biology and nonlinear molecular interactions that lead to the development of cortical structure. Likewise, partial least squares (PLS) regression analyses have been extensively used for the discovery of components of the transcriptome that are correlated with neuroimaging variability \cite{krishnan2011partial}. PLS methods are successful methods for dimensionality reduction and covariance estimation but they are still linear statistical methods.

Proposed framework, on the other hand, brings a perspective on generative modelling to imaging transcriptomics. The model does not simply learn direct statistical associations, but instead learns latent biological representations, which can be used to reconstruct the transcriptomic organization, and also to predict the vulnerability of the cortex to neurodegeneration. This distinction is relevant because the neurobiological systems are inherently hierarchical, nonlinear and spatially organized.

The incorporation of spatial regularization further differentiates the present framework from conventional transcriptomic mapping studies. Cortical degeneration unfolds across anatomically continuous systems shaped by cortical topology and network organization \cite{seeley2009neurodegenerative,raj2012network}. By integrating graph-based spatial smoothness constraints directly into the optimization objective, the proposed model incorporates biologically informed cortical continuity principles absent from most conventional imaging transcriptomics approaches.

Importantly, the goal of the present study was not solely to maximize predictive accuracy, but rather to develop a biologically interpretable computational framework capable of bridging molecular organization and macroscale neurodegeneration. The proposed approach therefore extends imaging transcriptomics beyond descriptive association analysis toward spatially-aware generative neurobiology. These findings suggest that generative representation learning may provide a promising future direction for multiscale computational neuroscience and neurodegenerative disease modeling.

\subsection{Limitations}

Several limitations of the present study should be considered when interpreting the findings. First, the Allen Human Brain Atlas (AHBA) is derived from only six postmortem donor brains \cite{hawrylycz2012anatomically}, which limits the ability to fully capture inter-individual variability in human cortical transcriptomic organization. Although AHBA remains the most widely used resource for imaging transcriptomics, the relatively small donor count introduces potential sampling biases and demographic limitations.

Second, the transcriptomic measurements used in this study represent static postmortem gene expression profiles rather than dynamic molecular processes evolving across disease progression. Neurodegenerative disorders such as Alzheimer's disease involve temporally evolving molecular cascades, inflammatory responses, and network-level pathological propagation \cite{jack2018nia}. Thus, obtaining static transcriptomic measurements is only a partial picture of the dynamic biological processes that drive disease progression.

Thirdly, the present framework dealt with the region-level cortical organization not the subject-level prediction. The resulting maps of neurodegeneration vulnerability are thus maps of spatial susceptibility for groups of individuals, instead of individual trajectories of disease. The present study also lacked the longitudinal imaging data which would enable a direct modelling of temporal neurodegenerative processes along disease stages.

An additional limitation involves the simplified spatial regularization framework used in this study. Cortical adjacency was modeled using a basic neighborhood graph, which does not fully capture the complexity of structural connectivity, functional interactions, or transsynaptic pathological spread across brain networks \cite{raj2012network}. More biologically realistic connectome-aware constraints may improve future modeling approaches.

Finally, the present work should be viewed as a preliminary proof-of-concept framework for spatially-aware generative neurobiology. While the observed predictive performance and latent organization are promising, further validation across larger datasets, additional disorders, and multimodal neuroimaging cohorts will be necessary to establish broader generalizability.

\subsection{Future Work}

Several future directions may substantially extend the present framework and further strengthen biologically informed computational neurodegenerative modeling. One important extension involves replacing the current spatial smoothness constraint with graph neural network (GNN) architectures capable of directly modeling cortical topology and inter-regional interactions \cite{bronstein2017geometric}. Graph-based neural systems may better capture distributed disease propagation across large-scale brain networks while preserving anatomically informed cortical organization.

Future studies should additionally incorporate structural and functional connectome constraints derived from diffusion MRI and resting-state functional MRI. Neurodegenerative pathology spreads preferentially across interconnected cortical systems \cite{seeley2009neurodegenerative,raj2012network}; therefore, integrating structural connectivity information may substantially improve biologically realistic disease modeling. Connectome-aware generative architectures could provide a more accurate representation of network-level neurodegenerative vulnerability.

The ongoing extension of the existing region-level approach to the subject-level prediction with individual multimodal MRI data is another direction of interest. Additional subject-specific measurements such as cortical thickness, volumetric, diffusion, and functional connectivity data in the longitudinal cohorts available in ADNI could help develop the ability to model a subject's neurodegenerative trajectory. These strategies may eventually help to estimate and predict disease risk and progression on an individual basis.

In addition, there are promising generative modelling approaches based on diffusion that could be incorporated into the present generative framework, e.g. \cite{ho2020denoising}. Diffusion models may give better latent space representation learning and possibly result in biologically plausible trajectories of cortical degeneration throughout the stages of disease. In the same way, a combination of multimodal omics (transcriptomics, proteomics, epigenomics, and metabolomics) could enable a more holistic picture of molecular vulnerability landscapes in neurodegenerative disease.

Another major future direction is the longitudinal modeling. The progression of Alzheimer's takes place over a period of many years across evolving cortical systems \cite{jack2018nia}. Integrating temporal disease trajectories into generative neurobiology frameworks may therefore provide substantially richer insight into the mechanisms governing neurodegenerative propagation.

Collectively, these future directions align closely with emerging priorities in computational neuroscience emphasizing multiscale, multimodal, and biologically interpretable artificial intelligence frameworks for brain disease modeling. The present study therefore provides an initial foundation for future spatially-aware generative neurobiology systems integrating molecular organization, network topology, and longitudinal disease progression.

\section{Conclusion}
\label{sect:conclu}
This study introduced a cross-scale spatially-aware generative framework for modeling relationships between cortical transcriptomic organization and regional neurodegenerative vulnerability in Alzheimer's disease. By integrating Allen Human Brain Atlas transcriptomic profiles with ADNI-derived cortical neurodegeneration maps, the proposed framework established a biologically grounded computational link between microscale molecular organization and macroscale cortical degeneration.

The model demonstrated strong predictive correspondence between learned latent transcriptomic programs and observed cortical neurodegeneration vulnerability patterns across 68 cortical regions. Importantly, the framework moved beyond conventional correlation-based imaging transcriptomics approaches by learning biologically structured latent representations through generative modeling and transcriptomic reconstruction objectives. Latent-space analyses further revealed organized low-dimensional molecular structure associated with distributed cortical vulnerability organization.

A central contribution of the present work involves the incorporation of spatially-aware neurobiological constraints into generative transcriptomic modeling. By integrating graph-based spatial regularization directly into the learning objective, the framework preserved biologically plausible cortical smoothness and anatomically coherent vulnerability organization. The resulting vulnerability maps closely recapitulated known Alzheimer's disease-related degeneration patterns involving entorhinal, temporal, and limbic cortical systems.

More broadly, this study highlights the potential of generative representation learning for computational neurobiology and imaging transcriptomics. The findings suggest that latent molecular programs embedded within cortical transcriptomic architecture contribute meaningfully to selective neurodegenerative vulnerability across distributed brain systems. Such approaches may ultimately help bridge longstanding gaps between molecular neuroscience, systems neuroscience, and large-scale neurodegenerative disease modeling.

Although the present work represents a preliminary proof-of-concept framework, it establishes a foundation for future multiscale computational neuroscience systems integrating transcriptomics, connectomics, multimodal neuroimaging, and longitudinal disease progression modeling. Spatially-aware generative neurobiology may therefore provide a promising future direction for understanding the molecular and systems-level mechanisms underlying neurodegenerative disorders.

\section{Acknowledgements}
We used large language models solely to assist in optimizing and formatting computational scripts and refining linguistic style of this manuscript. The authors retain full accountability for the final content. 

Data used in preparation of this article were obtained from the Alzheimer’s Disease Neuroimaging Initiative (ADNI) database (adni.loni.usc.edu). As such, the investigators within the ADNI contributed to the design and implementation of ADNIand/orprovideddatabutdidnotparticipateintheanalysis or writing ofthis report. AcompletelistingofADNIinvestigators can be found at: http://adni.loni.usc.edu/wp-content/uploads/how\_to\_apply/ADNI\_Acknowledgement\_List.pdf Data collection and sharing for the Alzheimer’s Disease Neuroimaging Initiative (ADNI) is funded by the National Institute on Aging (National Institutes of Health Grant U19AG024904).

\section{Conflicts of Interest} 
The authors declare no competing interests.

\section{Data Availability Statement} 
The ADNI Data is available at: https://adni.loni.usc.edu/data-samples/adni-data/. The code will be published on zenodo/github after publication.

\bibliographystyle{unsrtnat}
\bibliography{references}  %%% Uncomment this line and comment out the ``thebibliography'' section below to use the external .bib file (using bibtex) .




% 	\bibitem{keshet2016prediction}
% 	Keshet, Renato, Alina Maor, and George Kour.
% 	\newblock Prediction-Based, Prioritized Market-Share Insight Extraction.
% 	\newblock In {\em Advanced Data Mining and Applications (ADMA), 2016 12th International 
%                       Conference of}, pages 81--94,2016.

% \end{thebibliography}

\end{document}